\documentclass[aps,preprint]{revtex4}%
\usepackage{amsfonts}
\usepackage{amsmath}
\usepackage{amssymb}
\usepackage{graphicx}%
\providecommand{\U}[1]{\protect\rule{.1in}{.1in}}

\begin{document}
\preprint{HEP/123-qed}
\title[Short title for running header]{Integrable hydrodynamic equations for initial chiral currents and infinite
hydrodynamic chains from WZNW model and string model of WZNW type with
$SU(2),SO(3),SP(2),SU(\infty),SO(\infty),SP(\infty)$constant torsions}
\author{D.J. Cirilo-Lombardo}
\affiliation{Bogoliubov Laboratory of Theoretical Physics Joint Institute for Nuclear
Research, 141980, Dubna, Russian Federation}
\author{V. D. Gershun}
\affiliation{ITP, NSC Kharkov Institute of Physics and Technology, Kharkov, UA}
\author{}
\affiliation{}

\begin{abstract}
The WZNW and string models are considered in the terms of the initial and
invariant chiral currents assuming that the internal and external torsions
coincide (anticoincide) and they are the structure constants of the
$SU(n),SO(n),$ $SP(n)$ Lie algebras. These models are the auxiliary problems
in order to construct integrable equations of hydrodynamic type. It was shown
that the WZNW and string models in terms of invariant chiral currents are
integrable for the constant torsion associated with the structure constants of
the $SU(2),$ $SO(3),$ $SP(2)$ and SU(3) algebras only. The equation of motion
for the density of the first Casimir operator was obtained in the form of the
inviscid Burgers equation. The solution of this equation is presented through
the Lambert function. Also, a new equation of motion for the initial chiral
current was found.

The integrable infinite hydrodynamic chains obtained from the WZNW and string
models are given in terms of invariant chiral currents with the $SU(2)$,
$SO(3)$, $SP(2)$ and with $SU(\infty)$, $SO(\infty)$, $SP(\infty)$ constant
torsions. Also, the equations of motion for the density of any Casimir
operator and new infinite dimensional equations of hydrodynamic type for the
initial chiral currents through the symmetric structure constant of
$SU(\infty)$, $SO(\infty),$ $SP(\infty)$ algebras are obtained.

\end{abstract}
\volumeyear{year}
\volumenumber{number}
\issuenumber{number}
\eid{identifier}
\date[Date text]{date}
\received[Received text]{date}

\revised[Revised text]{date}

\accepted[Accepted text]{date}

\published[Published text]{date}

\startpage{101}
\endpage{102}
\maketitle
\tableofcontents

\section{Introduction}

The integrability of the two dimensional WZNW and string models is based on
the existence of the infinite number of the local and nonlocal currents and on
their charges. The $n$-dimensional WZNW model is described by the chiral left
$J_{A}^{L}=g^{-1}\partial_{A}g$ or the chiral right$J_{A}^{R}=\partial
_{A}g\;g^{-1}$ currents for arbitrary space-time dimension ${A=1,...n}$, where
$g$ is an element of the symmetry group of the model,$J_{A}=J_{A}^{\mu}t_{\mu
}$ and $t_{\mu}$ are the generators of the corresponding Lie algebra. These
chiral currents are related to left and right multiplication on the group
space. The two dimensional models ($A=0,1)$ have the additional chiral
currents
\begin{align*}
J_{\mu}^{L}(t,x)  &  =\frac{J_{0\mu}+\delta_{\mu\nu}J_{1}^{\nu}}{\sqrt{2}%
}=U_{\mu}\left(  x+t\right) \\
J_{\mu}^{R}(t,x)  &  =\frac{J_{0\mu}-\delta_{\mu\nu}J_{1}^{\nu}}{\sqrt{2}%
}=V_{\mu}\left(  x-t\right)
\end{align*}
related to the dynamic on the $(t,x)$ plane. The chiral currents $U_{\mu}$,
$V_{\mu}$ play an important role for a construction and study of the
integrable systems. In the $\sigma$-$\operatorname{mod}$el is not possible a
priori, to restrict the dynamics to only one mode (left or right). The only
possibility is to introduce a Witten term to the Wess-Zumino model. This term
introduces a potential for the torsion tensor over the curved space of the
group parameters in the addition to the metric tensor. Then, under some
conditions between the constant torsion and the structure constant of Lie
algebra the possibility to restrict the motion to one mode (left or right)
arises. String models with the antisymmetric background field was considered
in the conformal and light-cone gauges. This antisymmetric field plays the
role of potential of the external torsion in addition to the internal torsion,
that is directly related to the metric tensor. Under some conditions between
the both constant torsions, it is possible to introduce the chiral currents
$U_{\mu}$ and $V_{\mu}$ . Such currents form independent Kac-Moody algebras
making possible the restriction of the string dynamics to one mode. If we
assume that both torsions are precisely the structure constants of a Lie
algebra, the transverse coordinates of string belong to the compact space of
the corresponding Lie group. Consequently, under these assumptions, the WZNW
and string models with the antisymmetric field coincide. Supersymmetric models
and the integrability condition in supergroup spaces will be not treated here
we refer the reader to references\cite{Is},\cite{Is1},\cite{Is2} and\cite{Is3}
for a general description of such cases. The paper is organized as follows. In
Section 2 is devoted to the WZNW model: in subsection 2.1 the Lagrangian and
equation of motion are considered in the repere formalism obtaining the
antisymmetric field $B_{ab}$ in terms of the repere. The Hamiltonian formalism
and the commutations relations for new variables are focused in Subsections
2.2 and 2.3 : these variables are the chiral currents under the condition that
the external torsion coincides (anti-coincides) with the structure constants
of the $SU(n)$, $SO(n)$, $SP(n)$ algebras. In Section 3 we consider integrable
WZNW model with the constant torsion: in subsections 3.1, 3.2 \ the models
with the $SU(2)$, $SO(3)$, $SP(2)$ and $SU(3)$ constant torsion are
considered, the equation of motion for density of first Casimir operator is
obtained as the inviscid Burgers equation and its solution expressed as the
Lambert function. As a bonus a new nonhomogeneous equation for the initial
chiral current is obtained. The integrable infinite dimensional hydrodynamic
chains for WZNW model with the constant $SU(2)$, $SO(3)$, $SP(2)$ and
$SU(\infty)$, $SO(\infty)$, $SP(\infty)$ torsions are the subject of
subsection 3.3: new equations of motion of hydrodynamic type for the initial
chiral currents in terms of the symmetric structure constant of the
$SU(\infty)$, $SO(\infty)$, $SP(\infty)$ algebras are presented.

An integrable string model as auxiliary problem to integrable hydrodynamic
equations is treated in Section 4. In subsection 4.1 the string model in the
conformal and light-cone gauges is formulated obtaining the Hamiltonian and
the conditions on the metric and antisymmetric tensors. In subsection 4.2 we
show that the same Hamiltonian can be obtained from the flat string model in
the field theory approach. The subsection 4.3 we obtain and analyze the
commutation relations of the new variables, which are the chiral currents
under the condition that an internal and external torsions coincide
(anti-coincide) obeying the Kac-Moody algebras. In subsection 4.4 we presented
a short review of the string models of the hydrodynamic type for the case when
both torsions are null. In the short Section 5 a string model with constant
torsions is considered. The mathematical description of the equations of
motion for the invariant chiral and initial chiral currents of the string
model with the constant torsion in this section coincides with the
mathematical description of the equations of motion for the invariant chiral
and the initial chiral currents of the WZNW model with the constant torsion in
section 3. Finally some concluding remarks are given.

\section{Integrable WZNW model}

In this section we will show how the Poisson brackets (PBs) and the equations
of motion for the infinite hydrodynamic chains from the WZNW dynamic model are
obtained. Also, new infinite dimensional PBs and new nonlinear equations of
motion of the hydrodynamic type by using the symmetric structure constants of
$SU(\infty)$, $SO(\infty)$, $SP(\infty)$ Lie algebras are presented and
analyzed. \noindent

\subsection{Lagrangian and equations of motion}

The conformal invariant two-dimensional non-linear sigma model is described by
WZNW model which is nothing more that the sigma model \cite{Wei}-\cite{VZB}
with the Wess-Zumino term \cite{WZ}-\cite{Bus}, \cite{VZB} on the group
manifold. To each point of a dimensional world-sheet one associated an element
$g$ of a group $G$. We want to construct an action (Lagrangian density) with
the element of volume of the two-dimensional space invariant under of a group
transformations
\begin{equation}
S=\frac{1}{4}\int\frac{Tr(\omega\wedge dx^{\alpha})(\omega\wedge dx^{\beta
})\eta_{\alpha\beta}}{\epsilon_{\lambda\rho}dx^{\lambda}\wedge dx^{\rho}%
}+\frac{1}{2}\int Tr{(\omega(d)\wedge\omega(d)\wedge\omega(d))}. \label{eq1}%
\end{equation}
Here $x^{\alpha}=(t,x)$ are coordinates of the flat two-dimensional space:
$\alpha=(0,1)$ with signature $(-1,1)$ and $\eta_{\alpha\beta}$ is diagonal
metric of this space. The form $\omega(d)=\omega(d)^{\mu}t_{\mu}$ is the
differential Cartan one-form which belongs to a simple Lie algebra
\begin{equation}
\lbrack t_{\mu},t_{\nu}]=2iC_{\mu\nu}^{\lambda}t_{\lambda},\;Tr(t_{\mu}t_{\nu
})=2g_{\mu\nu},\;(\mu,\nu=1,2,...,n)\;. \label{eq2}%
\end{equation}
In any parametrization Cartan forms $\omega(d)=(g^{-1}dg)^{\mu}t_{\mu}$ depend
on the group parameters $\phi^{a}$: $\omega(d)=\omega(\phi,d\phi).$ The first
term of the Lagrangian (\ref{eq1}) has form
\[
\frac{Tr(\omega\wedge dx^{\alpha})(\omega\wedge dx^{\beta})\eta_{\alpha\beta}%
}{\epsilon_{\lambda\rho}dx^{\lambda}\wedge dx^{\rho}}=\frac{2}{d^{2}x}%
g_{\mu\nu}\omega_{a}^{\mu}\frac{\partial\phi^{a}}{\partial x^{\gamma}%
}(dx^{\gamma}\wedge dx^{\alpha})\omega_{b}^{\nu}\frac{\partial\phi^{b}%
}{\partial x^{\rho}}(dx^{\rho}\wedge dx^{\beta})\eta_{\alpha\beta}%
\]%
\begin{equation}
=2g_{ab}(\phi)\frac{\partial\phi^{a}}{\partial x^{\alpha}}\frac{\partial
\phi^{b}}{\partial x^{\beta}}\eta^{\alpha\beta}d^{2}x. \label{eq3}%
\end{equation}
Here we introduce notation
\begin{equation}
g_{ab}(\phi)=g_{\mu\nu}\omega_{a}^{\mu}\omega_{b}^{\nu},\;dx^{\gamma}\wedge
dx^{\alpha}=\epsilon^{\gamma\alpha}d^{2}x. \label{eq4}%
\end{equation}
One can see, that $g_{ab}(\phi)$ is metric tensor on the curve space of local
fields $\phi^{a},\;(a=1,2,...,n)\;.$ The $\omega_{a}^{\mu}(\phi)$ forms a
repere basis on the tangent space with the metric $g_{\mu\nu}$ in arbitrary
point of the curve space $\phi^{a}$. In the paper of \cite{Bra} the
antisymmetric field $B_{ab}(\phi)$ was obtained from the second term of action
(\ref{eq1}) for the metric tensor $g_{ab}=\delta_{ab}+(1-|\phi|^{2})^{-1}%
\phi_{a}\phi_{b},\,(a,b=1,2,3).$ Here we want to rewrite the WZNW model as
$\sigma$ - model of string type with antisymmetric field $B_{ab}(\phi)$ in
terms of the repere for the arbitrary metric $g_{ab}(\phi)$ and for any
dimension $n:$
\[
Tr{(\omega(d)\wedge\omega(d)\wedge\omega(d))}=2iC_{\mu\nu\lambda}\omega^{\mu
}(d)\wedge\omega^{\nu}(d)\wedge\omega^{\lambda}(d)=2iC_{\mu\nu\lambda}%
\Omega_{A}^{\mu}\Omega_{B}^{\nu}\Omega_{C}^{\lambda}\;dx^{A}\wedge
dx^{B}\wedge dx^{C}=
\]%
\[
=2g_{\mu\nu}\Omega_{A}^{\mu}\partial_{C}\Omega_{B}^{\nu}\epsilon^{ABC}%
d^{3}x=2g_{\mu\nu}\Omega_{a}^{\mu}\frac{\partial\Phi^{a}}{\partial x^{A}}%
\frac{\partial\Omega_{b}^{\nu}}{\partial x^{C}}\frac{\partial\Phi^{b}%
}{\partial x^{B}}\epsilon^{ABC}d^{3}x=
\]%
\begin{equation}
=g_{\mu\nu}(\Omega_{a}^{\mu}\frac{\partial\Omega_{b}^{\nu}}{\partial x^{C}%
}-\Omega_{b}^{\mu}\frac{\partial\Omega_{a}^{\nu}}{\partial x^{C}}%
)\frac{\partial\Phi^{a}}{\partial x^{A}}\frac{\partial\Phi^{b}}{\partial
x^{B}}\epsilon^{ABC}d^{3}x. \label{eq5}%
\end{equation}
In first line of (\ref{eq5}) the integrability condition of equation
$\partial_{A}g=g\Omega_{A}^{\mu}t_{\mu}$ was used
\[
\partial_{A}\Omega_{B}^{\mu}-\partial_{B}\Omega_{A}^{\mu}+2iC^{\mu\nu\lambda
}\Omega_{A}^{\nu}\Omega_{B}^{\lambda}=0.
\]
Here $x^{A}$ $(A=0,1,2)$ are coordinates of three dimension space-time,
$\Omega^{\mu}(d)$ is the associated one form on this space. Let us to separate
the second component of $A$ $(A=\alpha,2;\;\alpha=0,1)$ in the last line of
(\ref{eq5}). Then, the second term of the action takes the following
form:\noindent%
\begin{equation}
\int g_{\mu\nu}\epsilon^{\alpha\beta2}(\Omega_{a}^{\mu}\partial_{2}\Omega
_{b}^{\nu}-\partial_{2}\Omega_{a}^{\mu}\Omega_{b}^{\nu})\frac{\partial\Phi
^{a}}{\partial x^{\alpha}}\frac{\partial\Phi^{b}}{\partial x^{\beta}}%
d^{3}x=\int d^{2}x\int\limits_{0}^{M}\epsilon^{\alpha\beta2}B_{ab2}%
\frac{\partial\Phi^{a}}{\partial x^{\alpha}}\frac{\partial\Phi^{b}}{\partial
x^{\beta}}dx^{2} \label{eq6}%
\end{equation}
Here $B_{ab2}=g_{\mu\nu}(\Omega_{a}^{\mu}\partial_{2}\Omega_{b}^{\nu}%
-\partial_{2}\Omega_{a}^{\mu}\Omega_{b}^{\nu})=-B_{ba2}.$ The integration over
the coordinate $x^{2}$ was performed in the limits ( $0$ , $M$ ) with the
following boundary conditions:
\[
\Phi^{a}(x^{\alpha},x^{2})\left.  {}\right\vert ^{x^{2}=M}=\phi^{a}(x^{\alpha
}),B_{ab2}(x^{\alpha},x^{2})\left.  {}\right\vert ^{x^{2}=M}=B_{ab}(x^{\alpha
}).
\]
The integral in $x^{2}$ on the lower limit of integration equals zero, what is
easy to see by using the expansion of the integrand into the Taylor series.
Therefore, the second term of action becomes to:
\begin{equation}
\frac{1}{2}\int\epsilon^{\alpha\beta}B_{ab}\frac{\partial\phi^{a}}{\partial
x^{\alpha}}\frac{\partial\phi^{b}}{\partial x^{\beta}}d^{2}x. \label{eq7}%
\end{equation}
Consequently, the total action is:
\begin{equation}
S=\frac{1}{2}\int d^{2}x[g_{ab}(\phi)\eta^{\alpha\beta}+B_{ab}(\phi
)\epsilon^{\alpha\beta}]\frac{\partial{\phi^{a}}}{\partial{x^{\alpha}}}%
\frac{\partial{\phi^{b}}}{\partial{x^{\beta}}}. \label{eq8}%
\end{equation}
Here $g_{ab}(\phi)=g_{ba}(\phi)$ is the metric tensor of a group space $G$ and
$\phi^{a}(x)$ are the corresponding group parameters, $a,b=1,2,...n$. For
compact groups of dimension $n$, the space-time signature is $(n,0)$. The
background field $B_{ab}(\phi)$ on the group space $G$ is the antisymmetric
tensor field $B_{ab}(\phi(x))=-B_{ba}(\phi(x))$. The coordinates $x^{\alpha
}=(t,x),\;\alpha=0,1$ belong to a 2-dimensional word-sheet with the constant
metric tensor $\eta_{\alpha\beta}$ and signature $(-1,1)$. The second order
equation of motion has following form:
\begin{gather}
g_{ab}(\phi)\eta^{\alpha\beta}\frac{{\partial}^{2}\phi^{b}}{\partial
x^{\alpha}\partial x^{\beta}}+\Gamma_{abc}\eta^{\alpha\beta}\frac{\partial
\phi^{b}}{\partial x^{\alpha}}\frac{\partial\phi^{c}}{\partial x^{\beta}%
}+H_{abc}\epsilon^{\alpha\beta}\frac{\partial\phi^{b}}{\partial x^{\alpha}%
}\frac{\partial\phi^{c}}{\partial x^{\beta}}=0,\label{eq9}\\
\Gamma_{abc}=\frac{1}{2}(\frac{\partial g_{ab}}{\partial\phi^{c}}%
+\frac{\partial g_{ac}}{\partial\phi^{b}}-\frac{\partial g_{bc}}{\partial
\phi^{a}}),H_{abc}=\frac{\partial B_{ab}}{\partial\phi^{c}}+\frac{\partial
B_{ca}}{\partial\phi^{b}}+\frac{\partial B_{bc}}{\partial\phi^{a}}.
\label{eq10}%
\end{gather}
Here $\Gamma_{abc}(\phi)$ are the Christoffel symbols. It is a symmetric
function in $b,c$. The function $H_{abc}(\phi)$ is a total antisymmetric
function in $a$, $b$, $c$. If $H_{abc}(\phi)$ equals zero the antisymmetric
term in the Lagrangian is a pure topological one. The case $H_{abc}\neq0$
describes the WZNW model of string type. The antisymmetric field $B_{ab}$ in
equation (\ref{eq8}) in the light-cone variables $\frac{x\pm t}{\sqrt{2}}$ can
be considered as the antisymmetric part of a metric. It is the torsion
potential \cite{Bus}, \cite{Tse}. This form of the equation of the motion gave
the possibility to introduce symplectic and Poisson structures on the loop
spaces of smooth manifolds \cite{Mok6}, \cite{Mok5}. Let us introduce a repere
$e_{\mu}^{a}(\phi)=\omega_{\mu}^{a}$ on the compact group space $G$ and its
inverse $e_{a}^{\mu}(\phi)=\omega_{a}^{\mu}$ such that the metric tensor can
be written as
\begin{equation}
g_{ab}(\phi)=e_{a}^{\mu}(\phi)e_{b}^{\nu}(\phi)\delta_{\mu\nu},\;\delta
_{\mu\nu}=e_{\mu}^{a}(\phi)e_{\nu}^{b}(\phi)g_{ab}(\phi). \label{eq11}%
\end{equation}
Here $\delta_{\mu\nu}(\mu,\nu=1,2,...n)$ is a constant tensor on the tangent
space of the compact group space $G$ at some point $\phi^{a}(x)$ with the same
signature as $g_{ab}(\phi)$. To introduce the Hamiltonian we rewrite the
Lagrangian density and the equation of motion in terms of the world-sheet
coordinates $(t,x)$
\begin{equation}
L=\frac{1}{2}g_{ab}(\phi)[\frac{\partial\phi^{a}}{\partial t}\frac
{\partial\phi^{b}}{\partial t}-\frac{\partial\phi^{a}}{\partial x}%
\frac{\partial\phi^{b}}{\partial x}]+B_{ab}(\phi)\frac{\partial\phi^{a}%
}{\partial t}\frac{\partial\phi^{b}}{\partial x}. \label{eq12}%
\end{equation}
The equation of motion finally has the form:
\begin{equation}
g_{ab}(\phi)[\frac{\partial^{2}\phi^{a}}{{\partial t}{\partial t}}%
-\frac{\partial^{2}\phi^{a}}{{\partial x}{\partial x}}]+\Gamma_{abc}%
(\phi)[\frac{\partial\phi^{b}}{\partial t}\frac{\partial\phi^{c}}{\partial
t}-\frac{\partial\phi^{b}}{\partial x}\frac{\partial\phi^{c}}{\partial
x}]+2H_{abc}(\phi)\frac{\partial\phi^{b}}{\partial t}\frac{\partial\phi^{c}%
}{\partial x}=0. \label{eq13}%
\end{equation}

\subsection{Canonical momentum and Hamiltonian}

The canonical momentum is defined as
\begin{equation}
p_{a}(\phi(t,x))=\frac{\delta L}{\delta(\frac{\partial\phi^{a}}{\partial t}%
)}=g_{ab}(\phi)\frac{\partial\phi^{b}}{\partial t}+B_{ab}(\phi)\frac
{\partial\phi^{b}}{\partial x}. \label{eq14}%
\end{equation}
Consequently, the Hamiltonian has the following form:
\begin{equation}
H(\phi,p)=p_{a}\frac{\partial\phi^{a}}{\partial t}-L=\frac{1}{2}g^{ab}%
(\phi)[p_{a}-B_{ac}(\phi)\frac{\partial\phi^{c}}{\partial x}][p_{b}%
-B_{bd}(\phi)\frac{\partial\phi^{d}}{\partial x}]+\frac{1}{2}g_{ab}(\phi
)\frac{\partial\phi^{a}}{\partial x}\frac{\partial\phi^{b}}{\partial x}.
\label{eq15}%
\end{equation}
Let us introduce new dynamical variables
\begin{equation}
J_{0\mu}(\phi)=e_{\mu}^{a}(\phi)[p_{a}-B_{ab}(\phi)\frac{\partial\phi^{b}%
}{\partial x}],\;J_{1\mu}(\phi)=e_{a}^{\mu}(\phi)\frac{\partial\phi^{a}%
}{\partial x}. \label{eq16}%
\end{equation}
We see that Hamiltonian (\ref{eq15}) is factorized in these variables
\begin{equation}
H=\frac{1}{2}[\delta^{\mu\nu}J_{0\mu}(\phi)J_{0\nu}(\phi)+\delta_{\mu\nu}%
J_{1}^{\mu}(\phi)J_{1}^{\nu}(\phi)]. \label{eq17}%
\end{equation}
The equations of motion in terms of this variables are first order ones
\begin{equation}
\partial_{0}J_{1}^{\mu}(\phi)-\partial_{1}J_{0}^{\mu}(\phi)=C_{\nu\lambda
}^{\mu}J_{0}^{\nu}(\phi)J_{1}^{\lambda}(\phi),\;\;\partial_{0}J_{0}^{\mu}%
(\phi)-\partial_{1}J_{1}^{\mu}(\phi)=-H_{\nu\lambda}^{\mu}(\phi)J_{0}^{\nu
}(\phi)J_{1}^{\lambda}(\phi). \label{eq18}%
\end{equation}
Here $C^{\mu\nu\lambda}$ is the structure constant tensor which can be obtain
from the Maurer-Cartan equation:
\begin{equation}
C_{\nu\lambda}^{\mu}=\frac{\partial e_{a}^{\mu}(\phi)}{\partial x^{b}}[e_{\nu
}^{b}(\phi)e_{\lambda}^{a}(\phi)-e_{\nu}^{a}(\phi)e_{\lambda}^{b}%
(\phi)]=[\frac{\partial e_{a}^{\mu}(\phi)}{\partial x^{b}}-\frac{\partial
e_{b}^{\mu}(\phi)}{\partial x^{a}}]e_{\nu}^{b}(\phi)e_{\lambda}^{a}(\phi)
\label{eq19}%
\end{equation}
and similarly
\[
H_{\nu\lambda}^{\mu}(\phi)=g^{\mu\rho}H_{abc}(\phi)e_{\rho}^{a}(\phi)e_{\nu
}^{b}(\phi)e_{\lambda}^{c}(\phi).
\]

\subsection{Commutation relations for new variables}

The starting point is the canonical Poisson bracket (PB):
\begin{equation}
\{\phi^{a}(x),p_{b}(y)\}=\delta_{b}^{a}\delta(x-y). \label{eq20}%
\end{equation}
Let us consider the commutation relations for the functions $J_{0\mu}%
(\phi(x)),\;J_{1\mu}(\phi(x))=\delta_{\mu\nu}J_{1}^{\nu}(\phi(x))$ on the
phase space under the PB (\ref{eq20})
\[
\{J_{0\mu}(\phi(x)),\;J_{0\nu}(\phi(y))\}=C_{\mu\nu}^{\lambda}J_{0\lambda
}(\phi(x))\delta(x-y)+H_{\mu\nu}^{\lambda}(\phi(x))J_{1\lambda}(\phi
(x))\delta(x-y),
\]%
\[
\{J_{0\mu}(\phi(x)),\;J_{1\nu}(\phi(y))\}=C_{\mu\nu}^{\lambda}J_{1\lambda
}(\phi(x))\delta(x-y)+g_{\mu\nu}\frac{\partial}{\partial x}\delta(x-y),
\]%
\begin{equation}
\{J_{1\mu}(\phi(x)),\;J_{1\nu}(\phi(y))\}=0. \label{eq21}%
\end{equation}
Let us introduce the chiral variables
\begin{equation}
U_{\mu}=\frac{J_{0\mu}+\delta_{\mu\nu}J_{1}^{\nu}}{\sqrt{2}},\;\;V_{\mu}%
=\frac{J_{0\mu}-\delta_{\mu\nu}J_{1}^{\nu}}{\sqrt{2}}. \label{eq22}%
\end{equation}
The chiral variables $U_{\mu}(\phi(x)),\;V_{\mu}(\phi(x))$ satisfy to
following commutation relations:
\[
\{U_{\mu}(\phi(x)),U_{\nu}(\phi(y))\}=\frac{1}{2\sqrt{2}}[(3C_{\mu\nu
}^{\lambda}+H_{\mu\nu}^{\lambda}(\phi(x)))U_{\lambda}(\phi(x))-
\]%
\[
-(C_{\mu\nu}^{\lambda}+H_{\mu\nu}^{\lambda}(\phi(x)))V_{\lambda}%
(\phi(x))]\delta(x-y)+\delta_{\mu\nu}\partial_{x}\delta(x-y),
\]%
\begin{equation}
\{V_{\mu}(\phi(x)),V_{\nu}(\phi(y))\}=\frac{1}{2\sqrt{2}}[(3C_{\mu\nu
}^{\lambda}-H_{\mu\nu}^{\lambda}(\phi(x)))V_{\lambda}(\phi(x))- \label{eq23}%
\end{equation}%
\[
-(C_{\mu\nu}^{\lambda}-H_{\mu\nu}^{\lambda}(\phi(x)))U_{\lambda}%
(\phi(x))]\delta(x-y)-\delta_{\mu\nu}\partial_{x}\delta(x-y),
\]%
\[
\{U_{\mu}(\phi(x)),V_{\nu}(\phi(y))\}=\frac{1}{2\sqrt{2}}[(C_{\mu\nu}%
^{\lambda}+H_{\mu\nu}^{\lambda}(\phi(x)))U_{\lambda}(\phi(x))+(C_{\mu\nu
}^{\lambda}-H_{\mu\nu}^{\lambda}(\phi(x)))V_{\lambda}(\phi(x))]\delta(x-y).
\]
The commutation relations (\ref{eq21}), (\ref{eq23}) are not Poisson brackets
because the torsion $H_{\mu\nu}^{\lambda}(\phi)$ is not a smooth function.
These commutation relations form an algebra, if $H_{\mu\nu}^{\lambda}(\phi)$
is a \textit{constant} tensor. The interesting cases arise if $H_{\mu\nu
}^{\lambda}=\pm C_{\mu\nu}^{\lambda}.$ In the case $H_{\mu\nu}^{\lambda
}=-C_{\mu\nu}^{\lambda}$ the variables $U_{\mu}(\phi)$ form the closed
Kac-Moody algebra \cite{Kac}, \cite{Moo} for the right chiral currents
\begin{equation}
\{U_{\mu}(\phi(x)),\;U_{\nu}(\phi(y))\}_{2}=C_{\mu\nu}^{\lambda}U_{\lambda
}(\phi(x))\delta(x-y)+\delta_{\mu\nu}\partial_{x}\delta(x-y). \label{eq24}%
\end{equation}
Here we note the PB (\ref{eq24}) as $PB_{2}$. The remaining relations are not
essential.
\[
\{V_{\mu}(\phi(x)),\;V_{\nu}(\phi(y))\}=C_{\mu\nu}^{\lambda}(2V_{\lambda}%
(\phi(x))-U_{\lambda}(\phi(x)))\delta(x-y)-\delta_{\mu\nu}\partial_{x}%
\delta(x-y),
\]%
\[
\{U_{\mu}(\phi(x)).\;V_{\nu}(\phi(y))\}=C_{\mu\nu}^{\lambda}V_{\lambda}%
(\phi(x))\delta(x-y).
\]
In the case of $H_{\mu\nu}^{\lambda}=C_{\mu\nu}^{\lambda}$ variables $V_{\mu
}(\phi)$ form the closed Kac-Moody algebra for the left chiral currents
\begin{equation}
\{V_{\mu}(\phi(x)),\;V_{\nu}(\phi(y))\}=C_{\mu\nu}^{\lambda}V_{\lambda}%
(\phi(x))-\delta_{\mu\nu}\partial_{x}\delta(x-y). \label{eq25}%
\end{equation}
so, the remaining relations now are:
\[
\{U_{\mu}(\phi(x)),U_{\nu}(\phi(y))\}=C_{\mu\nu}^{\lambda}(2U_{\lambda}%
(\phi(x))-V_{\lambda}(\phi(x)))+\delta_{\mu\nu}\partial_{x}\delta(x-y),
\]%
\[
\{U_{\mu}(\phi(x)),\;V_{\nu}(\phi(y))\}=C_{\mu\nu}^{\lambda}U_{\lambda}%
(\phi(x))\delta(x-y).
\]
Let me note that Kac-Moody algebra has considered as a hidden symmetry of a
two-dimensional chiral models \cite{Zakh}, \cite{Dol}. In the 1983 one of the
authors (VDG) and Volkov, Tkach \cite{Non} considered the algebra of the
nonlocal charges of the $\sigma$-model in the context of itsintegrability. We
shown that the nonlocal charges form the enveloping algebra over the Kac-Moody
algebra. Let us rewrite equations of motion (\ref{eq18}) in terms of variables
$U_{\mu}(\phi(x))$, $V_{\mu}(\phi(x))$
\[
\partial_{-}U_{\mu}(\phi((t,x))=\frac{1}{2}(C_{\mu}^{\nu\lambda}+H_{\mu}%
^{\nu\lambda})V_{\nu}(\phi(t,x))U_{\lambda}(\phi(t,x)),
\]%
\begin{equation}
\partial_{+}V_{\mu}(\phi(t,x))=\frac{1}{2}(C_{\mu}^{\nu\lambda}-H_{\mu}%
^{\nu\lambda})U_{\nu}(\phi(t,x))V_{\lambda}(\phi(t,x)). \label{eq26}%
\end{equation}
In the case $C_{\mu\nu}^{\lambda}=-H_{\mu\nu}^{\lambda}$ the equation of
motion is
\begin{equation}
\partial_{-}U_{\mu}(\phi(t,x))=0,\;\;\partial_{+}V_{\mu}(\o (t,x))=C_{\mu
}^{\nu\lambda}U_{\nu}(\phi)V_{\lambda}(\phi). \label{eq27}%
\end{equation}

If $C_{\mu\nu}^{\lambda}=H_{\mu\nu}^{\lambda}$ the equation of motion is
\begin{equation}
\partial_{+}V_{\mu}(\phi(t,x))=0,\;\;\partial_{-}U_{\mu}(\phi(t,x))=C_{\mu
}^{\nu\lambda}V_{\nu}(\phi)U_{\lambda}(\phi). \label{eq28}%
\end{equation}
We see from the equations (\ref{eq24}) and (\ref{eq27}) that the chiral
currents $U_{\mu}$ form the closed system in the first case and from the
equations (\ref{eq25}), (\ref{eq28}) we see that the chiral currents $V_{\mu}$
also form the closed system. Here we introduced the notation:
\[
\partial_{+}=\frac{1}{\sqrt{2}}(\partial_{0}+\partial_{1}),\;\partial
_{-}=\frac{1}{\sqrt{2}}(\partial_{0}-\partial_{1}).
\]
The chiral currents $U^{\mu}(x)$ are the generators of translations in the
curved space of the fields $\phi^{a}(x)$
\[
\delta_{c}\phi^{a}(x)=\{\phi^{a}(x),c^{\mu}U_{\mu}(x)(\phi(x))\}=c^{\mu}%
e_{\mu}^{a}(\phi(x)))=c^{a}(\phi(x))).
\]
Simultaneously, they are the generators of the group transformations (with the
structure constants $C_{\lambda}^{\mu\nu})$ in the tangent space.

\section{Integrable WZNW model with constant torsion}

\noindent One of the ways to construct an integrable dynamical system is as
follows. We must to have a hierarchy of a Hamiltonians and to find a hierarchy
of Poisson brackets. This way is more simple, if the dynamical system have
some group structure. Let the torsion $C_{abc}$ to be the structure constants
of a Lie algebra. In the bi-Hamiltonian approach to the integrable string
models with the constant torsion we have considered the conserved primitive
chiral invariant currents (densities of the dynamical Casimir operators)
$C_{n}(U(x))$, as the local fields of the Riemannian manifold. The primitive
and non-primitive local charges of the invariant chiral currents form the
hierarchy of the new Hamiltonians. The primitive invariant currents are the
densities of the Casimir operators.The non primitive currents are functions of
the primitive ones. The commutation relations (\ref{eq24}) show that the
currents $U^{\mu}$ form the closed algebra. Therefore, we will consider PBs of
the right chiral currents $U^{\mu}$ and the Hamiltonians constructed only from
the right currents. The constant torsion will does not contributes to the
equation of motion, but it gives the possibility to introduce the group
structure and to introduce the symmetric structure constants. This paper was
stimulated by the papers \cite{Gol}, \cite{Are}, \cite{Eva}, \cite{Mou} about
the local conserved charges in two dimensional models. Evans, Hassan, MacKay,
Mountain \cite{Eva} constructed the local invariant chiral currents as the
polynomials of the initial chiral currents of the $SU(n)$, $SO(n)$, $SP(n)$
principal chiral models. Their paper was based on the paper of de Azcarraga,
Macfarlane, MacKay, Perez Bueno \cite{Az} about the tensor invariants for the
simple Lie algebras.

Let $t_{\mu}$ to be the generators of the $SU(n)$, $SO(n)$, $SP(n)$ Lie
algebras:
\begin{equation}
\lbrack t_{\mu},t_{\nu}]=2iC_{\mu\nu\lambda}t_{\lambda}. \label{eq29}%
\end{equation}
There are additional relations for generators of Lie algebra in the defining
matrix representation. There is following relation for the symmetric double
product of the generators of $SU(n)$ algebra:
\begin{equation}
\{t_{\mu},t_{\nu}\}=\frac{4}{n}\delta_{\mu\nu}+2d_{\mu\nu\lambda}t_{\lambda
},\,\,\mu=1,...,n^{2}-1. \label{eq30}%
\end{equation}
Here $d_{\mu\nu\lambda}$ is the totally symmetric structure constant tensor.
The Killing tensor $g_{\mu\nu}$ (\ref{eq2}) equals $\delta_{\mu\nu}$ for the
compact Lie algebras. The similar relation for the totally symmetric triple
product of the $SO(n)$ and $SP(n)$ algebras has the form:
\begin{equation}
t_{(\mu}t_{\nu}t_{\lambda)}=v_{\mu\nu\lambda}^{\rho}\,t_{\rho}. \label{eq31}%
\end{equation}
Here $v_{\mu\nu\lambda\rho}$ is the totally symmetric structure constant
tensor. The invariant chiral currents are the Liouville coordinates and they
can be constructed as the product of the invariant symmetric tensors
\[
d_{(\mu_{1}...\mu_{n})}=d_{({\mu_{1}\mu_{2}}}^{k_{1}}d_{\mu_{3}k_{1}}^{k_{2}%
}...d_{\mu_{n-1}\mu_{n})}^{k_{n-3}},\;d_{\mu_{1}\mu_{2}}=\delta_{\mu_{1}%
\mu_{2}}%
\]
for $SU(n)$ group and the initial chiral currents $U^{\mu}(\phi(x))$. It is so
called "d-family" of the invariant chiral currents \cite{Kle}, \cite{Ras},
\cite{Sud}:
\begin{equation}
C_{n}(U(\phi(x)))=d_{(\mu_{1}...\mu_{n})}U_{\mu_{1}}U_{\mu_{2}}...U_{\mu_{n}%
},\;C_{2}(U(\phi(x)))=\delta_{\mu\nu}U^{\mu}U^{\nu}. \label{eq32}%
\end{equation}
Any of these currents satisfy the equation of motion $\partial_{-}C_{n}%
(U(\phi(t,x)))=0$. The similar construction can be used for $SO(n)$, $SP(n)$
groups. The invariant chiral currents can be constructed as product of the
invariant symmetric constant tensor
\[
v_{(\mu_{1}...\mu_{2n})}=v_{(\mu_{1}\mu_{2}\mu_{3}}^{\nu_{1}}v_{\mu_{4}\mu
_{5}}^{\nu_{1}\nu_{2}}...v_{\mu_{2n-2}\mu_{2n-1}\mu_{2n})}^{\nu_{2n-3}%
},\;v_{\mu_{1}\mu_{2}}=\delta_{\mu_{1}\mu_{2}}.
\]
and the initial chiral currents $U^{\mu}$ as
\begin{equation}
C_{2n}(U(\phi(x)))=v_{\mu_{1}...\mu_{2n}}U^{\mu_{1}}...U^{\mu_{2n}}%
,\;C_{2}(U(\phi(x)))=\delta_{\mu_{1}\mu_{2}}U^{\mu_{1}}U^{\mu_{2}}.
\label{eq33}%
\end{equation}
The invariant chiral currents for $SU(2)$, $SO(3)$, $SP(2)$ have form:
\begin{equation}
C_{2n}=(C_{2})^{n} \label{eq34}%
\end{equation}
Another family of the invariant symmetric currents $J_{n}$ based on the
invariant symmetric chiral currents of simple Lie groups, is realized as the
symmetric trace of the $n$ product chiral currents $U(x)=t_{\mu}U^{\mu},$
$\mu=1,...,n^{2}-1$
\begin{equation}
J_{n}(U(\phi(x)))=SymTr(U...U). \label{eq35}%
\end{equation}
These invariant currents are polynomials of the product of the basic chiral
currents $C_{k},\;k=2,3,...,k$.
\[
J_{2}=2C_{2},\,\,J_{3}=2C_{3},\,\,J_{4}=2C_{4}+\frac{4}{n}C_{2}^{2}%
,\,\,J_{5}=2C_{5}+\frac{8}{n}C_{2}C_{3},
\]%
\[
J_{6}=2C_{6}+\frac{4}{n}C_{3}^{2}+\frac{8}{n}C_{2}C_{4}+\frac{8}{n^{2}}%
C_{2}^{3},
\]%
\[
J_{7}=2C_{7}+\frac{8}{n}C_{3}C_{4}+\frac{8}{n}C_{2}C_{5}+\frac{24}{n^{2}}%
C_{2}^{2}C_{3},
\]%
\[
J_{8}=2C_{8}+\frac{4}{n}C_{4}^{2}+\frac{8}{n}C_{3}C_{5}+\frac{8}{n}C_{2}%
C_{6}+\frac{24}{n^{2}}C_{2}C_{3}^{2}+\frac{24}{n^{2}}C_{2}^{2}C_{4}+\frac
{16}{n^{3}}C_{2}^{4}.
\]

Let us introduce the PB of hydrodynamic type for the chiral currents in the
Liouvlle form \cite{Dub1}
\begin{equation}
\{C_{m}(\phi)(x),C_{n}(\phi(y))\}=-W_{mn}(\phi(y))\frac{\partial}{\partial
y}\delta(y-x)+W_{nm}(\phi(x))\frac{\partial}{\partial x}\delta(x-y).
\label{eq36}%
\end{equation}
The asymmetric Hamiltonian function $W_{mn}(U(\phi(x)))$ for the finite
dimensional $SU(n)$, $SO(n)$, $SP(n)$ groups has the following form
\begin{equation}
W_{mn}(C(U(x)))=\frac{n-1}{m+n-2}\sum_{k}a_{k}C_{m+n-2,k}(U(x)),\;\sum
_{k=0}a_{k}=mn. \label{eq37}%
\end{equation}
This PB can be rewritten as the PB of the hydrodynamic type
\[
\{C_{m}(U(x)),C_{n}(U(y))\}=-\frac{n-1}{m+n-2}\sum_{k}a_{k}\frac
{dC_{m+n-2,k}(U((x)))}{dx}\delta(x-y)-
\]%
\begin{equation}
-\sum_{k}a_{k}C_{m+n-2,k}(U(x))\frac{\partial}{\partial x}\delta
(x-y),\sum_{k=0}a_{k}=mn. \label{eq38}%
\end{equation}
Here we used the following equalities \cite{Non}, \cite{Dub1} to perform the
PB of the hydrodynamic type:
\[
B(y)A(x)\frac{\partial}{\partial x}\delta(x-y)=B(y)A(y)\frac{\partial
}{\partial x}\delta(x-y)-B(y)\frac{\partial A(y)}{\partial y}\delta(x-y),
\]%
\[
\frac{\partial A(y)}{\partial y}\delta(x-y)+A(x)\frac{\partial}{\partial
x}\delta(x-y)=A(y)\frac{\partial}{\partial x}\delta(x-y),\;\;\frac{\partial
}{\partial x}\delta(x-y)=-\frac{\partial}{\partial y}\delta(y-x).
\]
Here the invariant total symmetric currents $C_{n,k},\;k=1,2...$ are new
currents which are polynomials of the product of the basic invariant currents
$C_{n_{1}}C_{n_{2}}...C_{n_{n}}$, $n_{1}+...+n_{n}=n$. They can be obtained by
the calculation of the total symmetric invariant currents $J_{n}$ using the
different replacements of the double product (\ref{eq30}) for the $SU(n)$
group and of the triple product (\ref{eq31}) for the $SO(n)$, $SP(n)$ groups
in the expressions for the invariant currents $J_{n}$.
\[
J_{6}=Tr[t(tt)(tt)t]=2C_{6}+\frac{4}{n}C_{3}^{2}+\frac{8}{n}C_{2}C_{4}%
+\frac{8}{n^{2}}C_{2}^{3},
\]%
\[
J_{6}=Tr[(tt)(tt)(tt)]=2C_{6,1}+\frac{12}{n}C_{2}C_{4}+\frac{8}{n^{2}}%
C_{2}^{3},
\]%
\begin{equation}
J_{7}=Tr[t(tt)t(tt)t]=2C_{7}+\frac{8}{n}C_{3}C_{4}+\frac{8}{n^{2}}C^{2}%
C_{5}+\frac{24}{n^{2}}C_{2}^{2}C_{3}, \label{eq39}%
\end{equation}%
\[
J_{7}=Tr[(tt)(tt)(tt)t]=2C_{7,1}+\frac{4}{n}C_{3}C_{4}+\frac{12}{n^{2}}%
C^{2}C_{5}+\frac{24}{n^{2}}C_{2}^{2}C_{3},
\]%
\[
J_{8}=Tr[t(tt)tt(tt)t]=2C_{8}+\frac{4}{n}C_{4}^{2}+\frac{8}{n}C_{3}C_{5}%
+\frac{8}{n}C_{2}C_{6}+\frac{24}{n^{2}}C_{2}C_{3}^{2}+\frac{24}{n^{2}}%
C_{2}^{2}C_{4}+\frac{16}{n^{3}}C_{2}^{4},
\]%
\[
J_{8}=Tr[(tt)(tt)t(tt)t]=2C_{8,1}+\frac{4}{n}C_{4}^{2}+\frac{4}{n}C_{3}%
C_{5}+\frac{24}{n^{2}}C_{2}C_{3}^{2}+\frac{12}{n}C_{2}C_{6}+\frac{24}{n^{2}%
}C_{2}^{2}C_{4}+\frac{16}{n^{3}}C_{2}^{4},
\]%
\[
J_{8}=Tr[(tt)(tt)(tt)(tt)]=2C_{8,2}+\frac{4}{n}C_{4}^{2}+\frac{16}{n}%
C_{2}C_{6,1}+\frac{32}{n^{2}}C_{2}^{2}C_{4}+\frac{16}{n^{3}}C_{2}^{4},
\]%
\[
J_{8}=Tr[t(tt)(tt)(tt)t]=2C_{8,3}+\frac{12}{n}C_{2}C_{6}+\frac{8}{n}C_{3}%
C_{5}+\frac{24}{n^{2}}C_{2}^{2}C_{4}+\frac{24}{n^{2}}C_{2}C_{3}^{2}+\frac
{16}{n}C_{2}^{4}.
\]
The new chiral currents $C_{n,k}(U)$ have the form:
\[
C_{6,1}=d_{\mu\nu}^{k}d_{\lambda\rho}^{l}d_{\sigma\varphi}^{n}d^{kln}U^{\mu
}U^{\nu}U^{\lambda}U^{\rho}U^{\sigma}U^{\varphi},
\]%
\[
C_{7,1}=d_{\mu\nu}^{k}d_{\lambda\rho}^{l}d_{\sigma\varphi}^{n}d_{\tau}%
^{nm}d^{klm}U^{\mu}U^{\nu}U^{\lambda}U^{\rho}U^{\sigma}U^{\varphi}U^{\tau},
\]%
\[
C_{8,1}=[d_{\mu\nu}^{k}d_{\lambda}^{kl}d_{\rho}^{ln}][d_{\sigma\varphi}%
^{m}][d_{\tau\theta}^{p}]d^{nmp}(U^{8})^{\mu\nu\lambda\rho\sigma\varphi
\tau\theta},
\]%
\[
C_{8,2}=[d_{\mu\nu}^{k}][d_{\lambda\rho}^{l}][d_{\sigma\varphi}^{n}%
][d_{\tau\theta}^{m}]d^{klp}d^{nmp}(U^{8})^{\mu\nu\lambda\rho\sigma\varphi
\tau\theta}.
\]%
\[
C_{8,3}=[d_{\mu\nu}^{k}d_{\lambda}^{kl}][d_{\rho\sigma}^{n}d_{\varphi}%
^{nm}][d_{\tau\theta}^{p}]d^{lmp}(U^{8})^{\mu\nu\lambda\rho\sigma\varphi
\tau\theta}.
\]
The following figures Fig1, Fig2 show the graphic images of the basic and the
"monster" invariant currents.
Here are only $l=n-1$ primitive invariant tensors for $SU(n)$ algebra,
$l=\frac{n-1}{2}$ for $SO(n)$ algebra and $l=\frac{n}{2}$ for $SP(n)$ algebra.
Higher invariant currents $C_{n}$ for $n\geq l+1$ are non-primitive currents
and they are polynomials of the primitive ones. By using the formula
(\ref{eq35}) we can obtain the expression for these polynomials from the
condition $J_{k}=0$ for $k>l$ for the generating function
\[
det(1-\lambda t_{\mu}U^{\mu})=\exp{Tr(ln(1-\lambda U))}=\exp{(-\sum
_{k=2}^{\infty}\frac{\lambda^{k}}{k}J_{k})}.
\]
The charges corresponding to the non-primitive chiral currents $C_{n}$ are not
Casimir operators. Consequently the WZNW model is not the integrable system
for the group symmetry of the finite rank $l\geq1$.

\bigskip%
\begin{figure}
[ptb]
\begin{center}
\includegraphics[
natheight=5.286600in,
natwidth=11.754500in,
height=3.0813in,
width=6.826in
]%
{N6MY1Q00.wmf}%
\end{center}
\end{figure}
%

\begin{figure}
[ptb]
\begin{center}
\includegraphics[
natheight=6.844100in,
natwidth=10.312000in,
height=3.4982in,
width=5.8825in
]%
{N6MULF01.wmf}%
\end{center}
\end{figure}

\subsection{Integrable WZNW models with $SU(2)$, $SO(3)$, $SP(2)$ constant
torsions}

There is only one primitive invariant tensor of $SU(2)$, $SO(3)$, $SP(2)$
algebras. The invariant non primitive tensors for $n\geq2$ are functions of
the primitive tensor. Let us introduce the local chiral currents based on the
invariant symmetric polynomials on the $SU(2)$, $SO(3)$, $SP(2)$ Lie groups:
\[
C_{2}(U)=\delta_{\mu\nu}U^{\mu}U^{\nu},C_{2n}(U)=(\delta_{\mu\nu}U^{\mu}%
U^{\nu})^{n},
\]
where $n=1,2,...$ and $\mu,\nu=1,2,3.$ The PB of Liouville coordinate
$C_{2}(U(x))$ has the following forms:
\[
\{C_{2}(U(x)),C_{2}(U(y))\}=-2C_{2}(U(y))\partial_{y}\delta(y-x)+2C_{2}%
(U(x))\partial_{x}\delta(x-y),
\]%
\[
\{C_{2}(U(x)),C_{2}(U(y))\}=4C_{2}(U(x))\partial_{x}\delta(x-y)+2\frac
{\partial}{\partial x}C_{2}(U(x))\delta(x-y).
\]
We will consider the chiral invariant $C_{2}(U(x))$ as a local field on the
Riemmann space of the chiral currents. As the Hamiltonians we choose the
following functions
\begin{equation}
H_{2(n+1)}=\frac{1}{2(n+1)}\int\limits_{0}^{2\pi}C_{2}^{n+1}%
(U(y))dy,\;n=0,1,...\infty. \label{eq40}%
\end{equation}
The equation of motion for the density of the first Casimir operator is as
follows
\begin{equation}
\frac{\partial C_{2}}{\partial t_{2(n+1)}}-(2n+1)(C_{2})^{n}\frac{dC_{2}}%
{dx}=0. \label{eq41}%
\end{equation}
The equation for the currents $C_{2}^{n}=C_{2n}$ is following:
\begin{equation}
\frac{\partial C_{2}^{n}}{\partial\tau_{n}}+(C_{2})^{n}\frac{dC_{2}^{n}}%
{dx}=0,\;\tau_{n}=-(2n+1)t_{2(n+1)}. \label{eq42}%
\end{equation}
This equation is the inviscid Burgers equation \cite{Bur}. We will find the
solution in the form:
\begin{equation}
C_{2}^{n}(\tau_{n},x)=\exp{(a+i(x-\tau_{n}C_{2}^{n}(\tau_{n},x)))}.
\label{eq43}%
\end{equation}
To obtain the solution of equation (\ref{eq42}) we rewrite this equation of
motion as follows:
\begin{equation}
Y_{n}=Z_{n}e^{Z_{n}},\;Y_{n}=i\tau_{n}e^{(a+ix)},\;Z_{n}=i\tau_{n}C_{2}^{n}.
\label{eq44}%
\end{equation}
The inverse transformation $Z_{n}=Z_{n}(Y_{n})$ is defined by the periodical
Lambert function \cite{Lam}:
\begin{equation}
C_{2}^{n}(\tau_{n},x)=\frac{1}{i\tau_{n}}W(i\tau_{n}e^{a+ix}). \label{eq45}%
\end{equation}
The solution for the first Casimir operator is consequently
\begin{equation}
C_{2}(t_{2(n+1)},x)=[\frac{i}{(2n+1)t_{2(n+1)}}W(-i(2n+1)t_{2(n+1)}%
e^{a+ix})]^{\frac{1}{n}}. \label{eq46}%
\end{equation}
The equation of motion for the initial chiral current $U^{\mu}$ defined by the
PB (\ref{eq24}) and the Hamiltonian (\ref{eq40})is
\begin{equation}
\frac{\partial U_{\mu}}{\partial t_{2(n+1)}}=\frac{\partial}{\partial
x}[U_{\mu}(UU)^{n}]=C_{2}^{n}\frac{\partial}{\partial x}U_{\mu}+U_{\mu}%
\frac{\partial}{\partial x}C_{2}^{n},\;=nU_{\mu}C_{2}^{n-1}\frac{\partial
}{\partial x}C_{2}+C_{2}^{n}\frac{\partial}{\partial x}U_{\mu},\;\mu=1,2,3.
\label{eq47}%
\end{equation}
It is easy to test, that equation of motion (\ref{eq47}) is in accordance with
equation (\ref{eq41}) by multiplication with the chiral current $U_{\mu}$ on
the both sides of equation (\ref{eq47}). It is possible to rewrite this
equation as the linear equation by using the solution (\ref{eq45}) which
diagonalize the equation (\ref{eq46})
\[
\frac{\partial U_{\mu}}{\partial t_{2(n+1)}}=\frac{\partial U_{\mu}}{\partial
x}f_{n}+U_{\mu}\frac{\partial}{\partial x}f_{n}%
\]
or as the linear nonhomogeneous equation
\begin{equation}
\frac{\partial z^{\mu}}{\partial t_{2(n+1)}}=f(t_{n},x)\frac{\partial z^{\mu}%
}{\partial x}+\frac{\partial}{\partial x}f(t_{n},x),\;\;z^{\mu}=\ln U^{\mu
},\;f=C_{2}^{n},\;\frac{\partial z^{\mu}}{\partial x}=\frac{1}{U^{\mu}}%
\frac{\partial U^{\mu}}{\partial x},\;(not\;sum). \label{eq48}%
\end{equation}

\subsection{Equation of motion for WZNW model with $SU(3)$ torsion}

The invariant chiral currents $C_{2}(U)$, $C_{3}(U)$ form a closed system. The
non-primitive currents have the following form
\[
C_{2n}=C_{2}^{n},\;C_{2n+1}=C_{2}^{n-1}C_{3}.
\]%
\[
C_{2}=\delta_{\mu\nu}U^{\mu}U^{\nu},\;\;C_{3}=d_{\mu\nu\lambda}U^{\mu}U^{\nu
}U^{\lambda},\;\mu\nu\lambda=1,2...8.
\]
The algebra of corresponding charges is a not abelian algebra, but the charges
$C_{2n}$ form the corresponding invariant subalgebra. The currents $C_{2}$ and
$C_{3}$ are the local coordinates on the Riemmann space and the invariant
currents $C_{2n}$ are densities of the Hamiltonians. Equation of motion for
$C_{3}$ is following:
\begin{equation}
\frac{\partial C_{3}(x)}{\partial t_{2(n+1)}}=-2C_{2}^{n}\frac{\partial
}{\partial x}C_{3}-6C_{3}\frac{\partial}{\partial x}C_{2}^{n}.. \label{eq49}%
\end{equation}
In terms of variables $g=lnC_{3}$, $f=C_{2}^{n}$ it is linear equation
\begin{equation}
\frac{\partial g}{\partial t_{2(n+1)}}+2f\frac{\partial}{\partial x}%
g+6\frac{\partial}{\partial x}f=0. \label{eq50}%
\end{equation}

\subsection{Infinite dimensional hydrodynamic chains}

The first example of the infinite dimensional hydrodynamic chains is based on
the invariant chiral currents $C_{2n}=(C_{2})^{n},\;n=1,2,...,\infty$ of the
WZNW model with the $SU(2)$, $SO(3)$, $SP(2)$ constant torsions. The PB of the
different degrees of the invariant chiral currents $C_{2}^{n}(x)$, $C_{2}%
^{m}(x)$ has form:
\begin{equation}
\{C_{2}^{m}(x),C_{2}^{n}(y)\}=\frac{2nm(2m-1)}{n+m-1}C_{2}^{n+m-1}%
(x)\frac{\partial\delta(x-y)}{\partial x}-\frac{2nm(2n-1)}{n+m-1}C_{2}%
^{n+m-1}(y)\frac{\partial\delta(y-x)}{\partial y}. \label{eq51}%
\end{equation}
The equation of motion for invariant current $C_{2}^{m}$ with Hamiltonian
\[
H_{2n}=\frac{1}{2n}\int\limits_{0}^{2pi}C_{2n}(y)dy
\]
has form:
\[
\frac{\partial C_{2}^{m}}{\partial t_{2n}}=\frac{m(2n-1)}{m+n-1}\frac{\partial
C_{2}^{m+n-1}}{\partial x}.
\]
After the redefinition $C_{2}^{n}=C_{2n}=C_{p}$ we can obtain the standard
form of the hydrodynamic chain
\begin{equation}
\{C_{p}(x),C_{q}(y)\}=\frac{pq(p-1)}{p+q-2}C_{p+q-2}(x)\frac{\partial
\delta(x-y)}{\partial x}-\frac{pq(q-1)}{p+q-2}C_{p+q-2}(y)\frac{\partial
\delta(y-x)}{\partial y}. \label{eq52}%
\end{equation}

The second example of the infinite dimensional chain is based on the invariant
chiral currents of the WZNW model with the $SU(\infty)$, $SO(\infty)$,
$SP(\infty)$ constant torsions. If the dimension of the matrix representation
$n$ is not ended $(n\rightarrow\infty)$, all the chiral currents are the
primitive currents. This is easy to see from the expression for the new chiral
currents $C_{m,k}$ . For example:
\[
C_{6,1}=C_{6}+\frac{2}{n}C_{3}^{2}-\frac{2}{n}C_{2}C_{4},\;\;C_{7,1}%
=C_{7}+\frac{4}{n}C_{3}C_{4}-\frac{4}{n}C_{2}C_{5},
\]%
\[
C_{8,1}=C_{8}+\frac{2}{n}C_{3}C_{5}-\frac{2}{n}C_{2}C_{6},\;\;C_{8,3}%
=C_{8}+\frac{2}{n}C_{4}^{2}-\frac{2}{n}C_{2}C_{6},
\]%
\[
C_{8,2}=C_{8}+\frac{4}{n}C_{3}C_{5}-\frac{4}{n}C_{2}C_{6}-\frac{4}{n^{2}}%
C_{2}C_{3}^{2}+\frac{4}{n^{2}}C_{2}^{2}C_{4}.
\]
The PB in Liouville coordinates $C_{m}(x),\;m=2,3,...,\infty$ has the form:
\begin{equation}
\{C_{m}(x),C_{n}(y)\}=-W_{mn}(C(y))\frac{\partial}{\partial y}\delta
(y-x)+W_{nm}(C(x))\frac{\partial}{\partial x}\delta(x-y), \label{eq53}%
\end{equation}%
\begin{equation}
W_{mn}(C(x))=\frac{mn(n-1)}{m+n-2}C_{m+n-2}(x). \label{eq54}%
\end{equation}
This PB satisfies the skew-symmetric condition $\{C_{m}(x),C_{n}%
(y)\}=-\{C_{n}(y),C_{m}(x)\}$. The Jacobi identity imposes conditions on the
Hamiltonian function $W_{mn}(C(x))$ \cite{Dub1}, \cite{Dub2}, \cite{Mal1}
\begin{equation}
(W_{kp}+W_{pk})\frac{\partial W_{mn}}{\partial C_{k}}=(W_{km}+W_{mk}%
)\frac{\partial W_{pn}}{\partial C_{k}},\;\frac{dW_{kp}}{dx}\frac{\partial
W_{nm}}{\partial C_{k}}=\frac{dW_{km}}{dx}\frac{\partial W_{np}}{\partial
C_{k}}. \label{eq55}%
\end{equation}
The Jacobi identity is satisfied by the metric tensor $W_{mn}(C(x))$
(\ref{eq54}) . The PB (\ref{eq53}) forms the algebra and can be rewritten as
the PB of the hydrodynamic type.
\begin{equation}
\{C_{m}(x),C_{n}(y)\}=\frac{mn(n-1)}{m+n-2}\frac{dC_{m+n-2}(x)}{dx}%
\delta(x-y)+mnC_{m+n-2}(x)\frac{\partial}{\partial x}\delta(x-y). \label{eq55}%
\end{equation}
The algebra of charges $\int\limits_{0}^{2\pi}C_{n}(x)dx$ is the abelian
algebra. Let us choose the Casimir operators $C_{n}$ as the Hamiltonians
\begin{equation}
H_{n}=\frac{1}{n}\int\limits_{0}^{2\pi}C_{n}(x)dx,\;n=2,3...\,. \label{eq56}%
\end{equation}
The equations of motion for the Casimir operator densities are the following
\begin{equation}
\frac{\partial C_{m}(x)}{\partial t_{n}}=\frac{1}{n}\int\limits_{0}^{2\pi
}[-W_{mn}(C(y))\frac{\partial}{\partial y}\delta(y-x)+W_{nm}C((x))\frac
{\partial}{\partial x}\delta(x-y)]dy=\frac{m(n-1)}{m+n-2}\frac{\partial
}{\partial x}C_{m+n-2}. \label{eq57}%
\end{equation}
Thus, the invariant chiral currents with the $SU(2)$, $SO(3)$, $SP(2)$
constant torsion and the invariant chiral currents with the $SU(\infty)$,
$SO(\infty)$, $SP(\infty)$ constant torsion form the same infinite
hydrodynamic chain (\ref{eq52}), (\ref{eq53}), ({\ref{eq57}). }

These PBs (\ref{eq53}), (\ref{eq56}) are particular case of the $M$-brackets
of Dorfman \cite{Dor} and Kupershmidt \cite{Kup1}, \cite{Kup2} for $M=2$ and
describe the hydrodynamic chains (see \cite{Pav1}, \cite{Pav2} and references
therein). We can construct the new nonlinear equations of motion for the
initial chiral currents $U^{\mu}$ using the flat $PB_{2}$ (\ref{eq24}) and
Hamiltonians $H_{n}$ (\ref{eq57}), where $C_{n}(x)$ defined by the equation
(\ref{eq32}) for $SU(\infty)$ group:
\[
\frac{\partial U_{\mu}(x)}{\partial t_{n}}=\frac{1}{n}\int\limits_{0}^{2\pi
}dy\{U_{\mu}(x),C_{n}(U(y))\}_{2},
\]%
\begin{equation}
\frac{\partial U_{\mu}(x)}{\partial t_{n}}=\frac{\partial}{\partial x}%
[d_{\nu_{1}\nu_{2}}^{k_{1}}d_{k_{1}\nu_{3}}^{k_{2}}...d_{\nu_{n-1}\mu
}^{k_{n-3}}U^{\nu_{1}}(x)...U^{\nu_{n-1}}(x)]. \label{eq59}%
\end{equation}
As an example we consider $n=3$:
\begin{equation}
\frac{\partial U_{\mu}}{\partial t_{3}}=\frac{\partial}{\partial x}(d_{\mu
\nu\lambda}U^{\nu}U^{\lambda}),\;\;\mu=1,2,...\infty. \label{eq60}%
\end{equation}
It is easy to see that this dynamical system is bi-Hamiltonian:
\begin{equation}
\frac{\partial U_{\mu}(x)}{\partial t_{3}}=\frac{1}{3}\int\limits_{0}^{2\pi
}dy\{U_{\mu}(x),C_{3}(U(y))\}_{2}=\frac{1}{2}\int\limits_{0}^{2\pi}dy\{U_{\mu
}(x),C_{2}(U(y))\}_{3}. \label{eq61}%
\end{equation}
Here $PB_{3}$ has form:
\begin{equation}
\{U_{\mu}(x),U_{\nu}(y)\}_{3}=2d_{\mu\nu\lambda}U^{\lambda}(x)\frac{\partial
}{{\partial x}}\delta(x-y). \label{eq62}%
\end{equation}
Let us remind that $d_{\mu\nu\lambda}$ are the symmetric structure constant of
the $SU(\infty)$ algebra in a matrix representation. This PB satisfies the
Jacobi identity for ($n\rightarrow\infty$)
\[
d_{\sigma\mu\nu}d_{\sigma\lambda\rho}+d_{\sigma\mu\lambda}d_{\sigma\nu\rho
}+d_{\sigma\mu\rho}d_{\sigma\nu\lambda}=\frac{1}{n}(\delta_{\mu\nu}%
\delta_{\lambda\rho}+\delta_{\mu\lambda}\delta_{\nu\rho}+\delta_{\nu\rho
}\delta_{\nu\lambda}).
\]

Similarly, we can obtain the equation of motion for the chiral currents of
$SO(\infty)$, $SP(\infty)$:
\begin{equation}
\frac{\partial U_{\mu}(x)}{\partial t_{n}}=\frac{\partial}{\partial x}%
[v_{\nu_{1}\nu_{2}\nu_{3}}^{k_{1}}...v_{\nu_{2n-2}\nu_{2n-1}\mu}^{k_{2n-3}%
}U^{\nu_{1}}...U^{\nu_{2n-1}}]. \label{eq63}%
\end{equation}
As an example we consider $n=4$:
\begin{equation}
\frac{\partial U_{\mu}}{\partial t_{4}}=\frac{\partial}{\partial x}(v_{\mu
\nu\lambda\rho}U^{\nu}U^{\lambda}U^{\rho}),\;\mu=1,2,...\infty. \label{eq64}%
\end{equation}
Also, we can obtain a solution for the metric function $W_{mn}(C(x))$ which is
analog to the Dubrovin-Novikov metric tensor $W_{\mu\nu}=\frac{\partial^{2}%
F}{\partial U^{\mu}\partial U^{\nu}}$ :
\[
C_{m}(U(x))=mF((U(x)),\;\;F(x,t_{n})=g(t_{n}+\frac{x}{n-1})
\]
and $g(t_{n}+\frac{x}{n-1})$ is an arbitrary function of its argument.

\section{Integrable string model}

In this section we will show that the description of the dynamics of the
transverse coordinates of the string in the conformal and light-cone gauges
coincides with the description of the dynamics of compact coordinates of WZNW
model (see \cite{Ger9}). Consequently,we will show here that the string model
can be considered as an auxiliary problem in order to obtain integrable
equations of hydrodynamic type.

\subsection{Lagrangian in the conformal and light-cone gauges}

The conformal invariant closed string model in the background gravity and
antisymmetric fields is described by the following action \cite{Gre} (see also
\cite{Rud} and references therein)
\begin{equation}
S=\frac{1}{2}\int\limits_{0}^{2\pi}d^{2}x\sqrt{g}[g^{\alpha\beta}%
g_{AB}(X)\frac{\partial X^{A}}{\partial x^{\alpha}}\frac{\partial X^{B}%
}{\partial x^{\beta}}+\frac{\epsilon^{\alpha\beta}}{\sqrt{g}}B_{AB}%
(X)\frac{\partial X^{A}}{\partial x^{\alpha}}\frac{\partial X^{B}}{\partial
x^{\beta}}].\label{eq65}%
\end{equation}
The action $S[g_{\alpha\beta},X^{A}]$ is a functional of the worldsheet metric
$g_{\alpha\beta}(x)$ and of the $n+2$ spacetime coordinate fields
$X^{A}(x)=X^{A}(x+2\pi)$. Here $x^{\alpha}$ $(\alpha=0,1)$ are world-sheet
coordinates, $X^{A}(x)$ $(A=0,a,n+1)$ are target coordinates and
$a=1,2,...,n.$ The signature of worldsheet space is $(-,+)$ and signature of
target space is $(-,+,...,+)$. Here $g$ is the determinant world-sheet metric
tensor $g_{\alpha\beta}$. The target space-time fields consist of the
symmetric metric tensor $g_{AB}(X)=g_{BA}(X)$ and antisymmetric tensor field
$B_{AB}(X)=-B_{BA}(X)$.The antisymmetric tensor $\epsilon^{\alpha\beta}$ is
the Levi-Civita tensor such that $\epsilon^{01}=1$. Notice that target
space-time indices are Roman, worldsheet indices are Greek and Einstein
convention is assummed. The world-sheet metric $g^{\alpha\beta}$ is an
auxiliary field in the Lagrangian (\ref{eq65}). The variation of Lagrangian
(\ref{eq65}) with respect to the field $g^{\alpha\beta}$ yields the classical
field equation:
\begin{equation}
g_{AB}(X)\frac{\partial X^{A}}{\partial x^{\alpha}}\frac{\partial X^{B}%
}{\partial x^{\beta}}-g_{\alpha\beta}g^{\gamma\delta}\frac{\partial X^{A}%
}{\partial x^{\gamma}}\frac{\partial X^{B}}{\partial x^{\delta}}%
=0.\label{eq66}%
\end{equation}
The Lagrangian (\ref{eq65}) is invariant under re-parametrization of the
world-sheet coordinates
\[
x^{\alpha}\rightarrow x^{\alpha}+\epsilon^{\alpha}(t,x).
\]
These transformations permit to consider the conformal gauge for the
world-sheet metric
\begin{equation}
g^{\alpha\beta}=e^{\theta(t,x)}\eta^{\alpha\beta}.\label{eq67}%
\end{equation}
and the light-cone gauge for the light-cone variable $X^{+}$: $X^{+}=p^{+}t.$
There $\eta^{\alpha\beta}$ is the diagonal metric of the flat world-sheet
$\eta^{\alpha\beta}=\{\eta^{00},\eta^{11}\}=(-1,1)$ and
\begin{equation}
X^{\pm}=\frac{X^{n+1}\pm X^{0}}{\sqrt{2}}.\label{eq68}%
\end{equation}
The Lagrangian (\ref{eq65}) does not depend of the conformal field
$\theta(t,x)$. The equation (\ref{eq66}) leads to the following relations:
\begin{equation}
g_{AB}(X)(\partial_{t}X^{A}\partial_{t}X^{B}+\partial_{x}X^{A}\partial
_{x}X^{B})=0,\;g_{AB}(X)\partial_{t}X^{A}\partial_{x}X^{B}=0.\label{eq69}%
\end{equation}
These relations describe the geodesic motion of the probe body on the two
dimensional surface, which is embedded in the $n+2$ curve target space. To
introduce the Hamiltonian, let us rewrite the Lagrangian density (\ref{eq65})
in terms of the world-sheet coordinates (t,x) in the conformal gauge
(\ref{eq67}). It has form:
\begin{equation}
L=\frac{1}{2}g_{AB}(X)[\frac{\partial X^{A}}{\partial t}\frac{\partial X^{B}%
}{\partial t}-\frac{\partial X^{A}}{\partial x}\frac{\partial X^{B}}{\partial
x}]+B_{AB}(X)\frac{\partial X^{A}}{\partial t}\frac{\partial X^{B}}{\partial
x}.\label{eq70}%
\end{equation}
The canonical momentum is the following
\begin{equation}
p_{A}(t,x)=\frac{\delta L}{\delta(\frac{\partial X^{A}}{\partial t})}%
=g_{AB}(X)\frac{\partial X^{B}}{\partial t}+B_{AB}(X)\frac{\partial X^{B}%
}{\partial x}.\label{eq71}%
\end{equation}
By the definition, the Hamiltonian has the following form:
\begin{equation}
H(X,p)=p_{A}\frac{\partial X^{A}}{\partial t}-L=\frac{1}{2}g^{AB}[p_{A}%
-B_{AC}(X)\frac{\partial X^{C}}{\partial x}][p_{B}-B_{BD}(X)\frac{\partial
X^{D}}{\partial x}]+\frac{1}{2}g_{AB}\frac{\partial X^{A}}{\partial x}%
\frac{\partial X^{B}}{\partial x}.\label{eq72}%
\end{equation}
Let us consider the constraints (\ref{eq69}), the equation for canonical
momentum (\ref{eq71}) and the definition of the inverse metric tensor
$g_{AB}\;g^{BC}=\delta_{A}^{C}$ in the light-cone gauge $X^{+}=p^{+}t$ to
delete nonphysical components of the metric tensor $g_{AB}$. From the
definition of the components of the momentum $p_{-}=-p^{+}$, we obtain the
following constraints:
\begin{equation}
g_{--}=g_{-a}=B_{--}=B_{-a}=g^{++}=g^{+a}=B^{++}=B^{+a}=0,g_{-+}%
=-1.\label{eq73}%
\end{equation}
The momentum component $p_{+}$ leads to constraints:
\begin{equation}
p_{+}=-p^{-}=-X^{-},\,B_{+-}=B_{-a}=0..\label{eq74}%
\end{equation}
The component $p_{a}$ has form
\begin{equation}
p_{a}=g_{ab}(X){\partial_{t}X_{a}}+B_{ab}(X){\partial_{x}X^{b}}.\label{eq75}%
\end{equation}
From the inverse metric we obtain the following constraints:
\begin{equation}
g^{+b}=-g_{-a}g^{ab}=0,\;g^{++}=g_{--}=0,\;g_{+-}=g^{-+}=-1.\label{eq76}%
\end{equation}
From the relations $g_{+A}g^{Ab}=0$ and $g_{+A}g^{A-}=0$ we obtained
constraints $g_{+a}=g_{++}=0$. These constraints were notobtained in the paper
\cite{Rud}. From the constraints $g_{AB}{\partial X_{t}^{A}}{\partial_{x}%
X^{B}}=0$ we obtain the relation
\begin{equation}
\partial_{x}X^{-}=\frac{1}{p^{+}}g_{ab}(X){\partial_{t}X^{a}}{\partial
_{x}X^{b}}.\label{eq77}%
\end{equation}
From the constraints
\[
g_{AB}({\partial_{t}X^{A}}{\partial_{t}X^{B}}+{\partial_{x}X^{A}}{\partial
_{x}X^{B}})=0
\]
the equation
\begin{equation}
{\partial_{t}X^{-}}=\frac{1}{2p^{+}}g_{ab}(X)({\partial_{t}X^{a}}{\partial
_{t}X^{b}}+{\partial_{x}X^{a}}{\partial_{x}X^{b}})\label{eq78}%
\end{equation}
is obtained. By using the equations (\ref{eq73}), (\ref{eq74}) and
(\ref{eq76}) we can write the Hamiltonian in the light-cone gauge:
\begin{equation}
H=p^{-}p^{+}=\frac{1}{2}[g^{ab}(p_{a}-B_{ac}{\partial_{x}X^{c}})(p_{b}%
-B_{bd}{\partial_{x}X^{d}})+g_{ab}{\partial_{x}X^{a}}{\partial_{x}X^{b}%
}].\label{eq79}%
\end{equation}
This Hamiltonian describe the transverse coordinates of string and it differ
from the Hamiltonian in paper \cite{Rud} due the constraints $g^{--}=g_{++}%
=0$,\thinspace$g^{-a}=-g_{+}^{a}=0$,\thinspace$B_{a}^{-}=-B_{+a}=0$. The
motion of the longitudinal coordinate $X^{-}(x,t)$ is described consequently
by the equations (\ref{eq77}), (\ref{eq78}).

\subsection{String Lagrangian of field theory type}

Historically, a classical field theory consider models of free fields without
interaction, study the group symmetry of them and representations of this
group of symmetry. Further, a theory of interactions is constructed for a
chosen representation with conservation of the corresponding group symmetry.
We apply this method for constructing a string model with transverse
coordinates. The action of the free string has form:
\begin{equation}
S=\frac{1}{2}\int\limits_{0}^{2\pi}d^{2}x\sqrt{g}g^{\alpha\beta}g_{AB}%
\partial_{\alpha}X^{A}\partial_{\beta}X^{B}. \label{eq80}%
\end{equation}
The difference this action from action (\ref{eq65}) consist of the lack of the
antisymmetric field $B_{AB}$ and by the constant metric tensor $g_{AB}$. In
the conformal gauge (\ref{eq67}), the Lagrangian has form:
\begin{equation}
L=\frac{1}{2}g_{AB}[\frac{\partial X^{A}}{\partial t}\frac{\partial X^{B}%
}{\partial t}-\frac{\partial X^{A}}{\partial x}\frac{\partial X^{B}}{\partial
x}] \label{eq81}%
\end{equation}
The canonical momentum is the following
\begin{equation}
p_{A}(X(t,x))=\frac{\delta L}{\delta(\frac{\partial X^{A}}{\partial t}%
)}=g_{AB}\frac{\partial X^{B}}{\partial t}. \label{eq82}%
\end{equation}
By the definition, the Hamiltonian has following form:
\begin{equation}
H(X,p)=p_{a}\frac{\partial X^{A}}{\partial t}-L=\frac{1}{2}g^{AB}p_{A}%
p_{B}+\frac{1}{2}g_{AB}\frac{\partial X^{A}}{\partial x}\frac{\partial X^{B}%
}{\partial x} \label{eq83}%
\end{equation}
The variation of the Lagrangian (\ref{eq80}) with respect to the field
$g^{\alpha\beta}$ yields two constraints:
\begin{equation}
g_{AB}(\partial_{t}X^{A}\partial_{t}X^{B}+\partial_{x}X^{A}\partial_{x}%
X^{B})=0,\;g_{AB}\partial_{t}X^{A}\partial_{x}X^{B}=0. \label{eq84}%
\end{equation}
Let us consider the constraints (\ref{eq84}) in the light-cone gauge
$X^{+}=p^{+}t$. Let us remember, that the target space fields $X^{A}$
($A=0,a,n+1)$ in the light-cone coordinates $(A=+,-,a)$ have form
(\ref{eq68})
\[
X^{A}=(X^{+},X^{-},X^{a})=(\frac{X^{n+1}+X^{0}}{\sqrt{2}},\frac{X^{n+1}-X^{0}%
}{\sqrt{2}},X^{a}).
\]
From the definition of the canonical momentum (\ref{eq82}) and of the inverse
metric tensor we obtain constraints for the constant metric
\[
g_{--}=g_{++}=g_{a+}=g_{a-}=g^{++}=g^{--}g^{a+}=g^{a-}=0,\;g_{+-}=g^{-+}=-1.
\]
The constraints (\ref{eq84}) leads to the following equations:
\begin{equation}
\partial_{x}X^{-}=\frac{1}{p^{+}}g_{ab}{\partial_{t}X^{a}}{\partial_{x}X^{b}}.
\label{eq85}%
\end{equation}%
\begin{equation}
{\partial_{t}X^{-}}=\frac{1}{2p^{+}}g_{ab}({\partial_{t}X^{a}}{\partial
_{t}X^{b}}+{\partial_{x}X^{a}}{\partial_{x}X^{b}}). \label{eq86}%
\end{equation}
By using the definition of the momentum (\ref{eq82}), we can write the
Hamiltonian in the light-cone gauge
\begin{equation}
H=p^{-}p^{+}=\frac{1}{2}[g^{ab}p_{a}p_{b}+g_{ab}{\partial_{x}X^{a}}%
{\partial_{x}X^{b}}]. \label{eq87}%
\end{equation}

Thus, we obtained the Hamiltonian which describe the transverse coordinates of
the flat target space. We want to obtain the Hamiltonian which describe the
interaction of the transverse coordinates. To do this, we will consider a
curved target space by the replacement $g_{ab}\rightarrow g_{ab}(X)$. An
interaction with the gauge field $B_{ab}(X){\partial_{x}X^{b}}$ can be
introduced to the Hamiltonian by minimal replacement $p_{a}\rightarrow
p_{a}-B_{ab}{\partial_{x}X^{b}}$. Thus, both methods leads to the same
Hamiltonian for the transverse target space coordinates. The moving of the
longitudinal coordinate is describe by the equations (\ref{eq77}), (\ref{eq78}).

\subsection{Repere formalism and commutation relations}

Let us return to the subsection (4.1). In the repere formalism
\[
g_{ab}(X)=e_{a}^{\mu}(X)e_{b}^{\nu}(X)g_{\mu\nu},
\]
where $g_{\mu\nu}$ is constant tensor of the tangent space tothe curved space
of the string coordinates $X^{a}$. The repere $e_{a}^{\mu}$ satisfies to the
condition
\[
g^{\mu\nu}=e_{a}^{\mu}(X)e_{b}^{\nu}(X)g^{ab}(X).
\]
To factorize the Hamiltonian (\ref{eq72}) we introduce new variables:
\begin{equation}
J_{0\mu}(X)=e_{\mu}^{a}(X)[p_{a}-B_{ab}(X)X^{^{\prime}b}],\;J_{1\mu}%
(X)=g_{\mu\nu}e_{a}^{\nu}{\partial_{x}X^{a}}. \label{eq88}%
\end{equation}
We see, that the Hamiltonian (\ref{eq72}) is factorized in this variables:
\begin{equation}
H=\frac{1}{2}[g^{\mu\nu}J_{0\mu}(X)J_{0\nu}(X)+g_{\mu\nu}J_{1}^{\mu}%
(X)J_{1}^{\nu}(X)] \label{eq89}%
\end{equation}
The canonical PB is the following:
\[
\{X^{a}(x),\;p_{b}(y)\}=\delta_{b}^{a}\delta(x-y).
\]
The commutation relations of new variables $J_{0\mu}(X(x)),\;J_{1\mu}(X(x))$
under PB have the following form:
\[
\{J_{0\mu}(X(x)),\;J_{0\nu}(X(y))\}=C_{\mu\nu}^{\lambda}(X(x))J_{0\lambda
}(x)\delta(x-y)+H_{\mu\nu}^{\lambda}(X(x))J_{1\lambda}(X(x))\delta(x-y),
\]%
\[
\{J_{0\mu}(X(x)),\;J_{1\nu}(X(y))\}=C_{\mu\nu}^{\lambda}(X(x))J_{1\lambda
}(X(x))\delta(x-y)+g_{\mu\nu}\frac{\partial}{\partial x}\delta(x-y),
\]%
\begin{equation}
\{J_{1\mu}(X(x)),\;J_{1\nu}(X(y))\}=0. \label{eq90}%
\end{equation}
Here the tensor $C_{\mu\nu}^{\lambda}(X)$ is the torsion related to the metric
tensor:
\begin{equation}
C_{\nu\lambda}^{\mu}(X)=\frac{\partial e_{a}^{\mu}}{\partial x^{b}}(e_{\nu
}^{b}e_{\lambda}^{a}-e_{\nu}^{a}e_{\lambda}^{b})=(\frac{\partial e_{a}^{\mu}%
}{\partial x^{b}}-\frac{\partial e_{b}^{\mu}}{\partial x^{a}})e_{\nu}%
^{b}e_{\lambda}^{a} \label{eq91}%
\end{equation}
The tensor $H_{\mu\nu}^{\lambda}(X)$ is the torsion related to the background
field $B_{ab}$:
\[
H_{\nu\lambda}^{\mu}(X)=G^{\mu\rho}H_{abc}(X)e_{\rho}^{a}(X)e_{\nu}%
^{b}(X)e_{\lambda}^{c}(X),
\]%
\begin{equation}
H_{abc}(X)=\frac{\partial B_{ab}}{\partial X^{c}}+\frac{\partial B_{ca}%
}{\partial X^{b}}+\frac{\partial B_{bc}}{\partial X^{a}}. \label{eq92}%
\end{equation}
The difference between the commutation relations (\ref{eq88}) of the string
model with the commutation relations of the WZNW model (\ref{eq21}) is due the
different tensors $C_{\mu\nu}^{\lambda}$. The constant torsion $C_{\mu\nu
}^{\lambda}$ of the WZNW model is the structure constant of the Lie algebra by
simple definition. The torsion $C_{\mu\nu}^{\lambda}(X(x))$ is the function of
the Riemmann space coordinates. In some applications this torsion may be zero
tensor or the constant tensor. Let us introduce the chiral variables:
\begin{equation}
U_{\mu}(X)=\frac{J_{0\mu}(X)+g_{\mu\nu}J_{1}^{\nu}(X)}{\sqrt{2}},\;\;V_{\mu
}(X)=\frac{J_{0\mu}(X)-g_{\mu\nu}J_{1}^{\nu}(X)}{\sqrt{2}}. \label{eq93}%
\end{equation}
The chiral variables $U_{\mu}(X(x)),\;V_{\mu}(X(x))$ satisfy the following
commutation relations:
\[
\{U_{\mu}(X(x)),U_{\nu}(X(y))\}=\frac{1}{2\sqrt{2}}[(3C_{\mu\nu}^{\lambda
}(X(x))+H_{\mu\nu}^{\lambda}(X(x)))U_{\lambda}(X(x))-
\]%
\[
-(C_{\mu\nu}^{\lambda}(X(x))+H_{\mu\nu}^{\lambda}(X(x)))V_{\lambda
}(X(x))]\delta(x-y)+g_{\mu\nu}\partial_{x}\delta(x-y),
\]%
\begin{equation}
\{V_{\mu}(X(x)),V_{\nu}(X(y))\}=\frac{1}{2\sqrt{2}}[(3C_{\mu\nu}^{\lambda
}(X(x))-H_{\mu\nu}^{\lambda}(X(x)))V_{\lambda}(X(x))- \label{eq94}%
\end{equation}%
\[
-(C_{\mu\nu}^{\lambda}(X(x))-H_{\mu\nu}^{\lambda}(X(x)))U_{\lambda
}(X(x))]\delta(x-y)-g_{\mu\nu}\partial_{x}\delta(x-y),
\]%
\[
\{U_{\mu}(X(x)),V_{\nu}(X(y))\}=\frac{1}{2\sqrt{2}}[(C_{\mu\nu}^{\lambda
}(X(x))+H_{\mu\nu}^{\lambda}(X(x)))U_{\lambda}(X(x))+
\]%
\[
+(C_{\mu\nu}^{\lambda}(X(x))-H_{\mu\nu}^{\lambda}(X(x)))V_{\lambda
}(X(x))]\delta(x-y).
\]
This commutation relations form an algebra, if tensors $C_{\lambda}^{\mu\nu
}(X)$, $H_{\mu\nu}^{\lambda}(X)$ are the constant tensors. The interesting
cases, again arise if $H_{\mu\nu}^{\lambda}=\pm C_{\mu\nu}^{\lambda}$ and
tensor $C_{\lambda}^{\mu\nu}$ is the structure tensor of the compact Lie
algebra. In the case $H_{\mu\nu}^{\lambda}=-C_{\mu\nu}^{\lambda}$ variables
$U_{\mu}$ form the closed Kac-Moody algebra for the right chiral currents
\begin{equation}
\{U_{\mu}(X(x)),\;U_{\nu}(X(y))\}=C_{\mu\nu}^{\lambda}U_{\lambda}%
(X(x))\delta(x-y)+\delta_{\mu\nu}\partial_{x}\delta(x-y). \label{eq95}%
\end{equation}
In the case $H_{\mu\nu}^{\lambda}=C_{\mu\nu}^{\lambda}$ variables $V_{\mu}$
form the closed Kac-Moody algebra for the left chiral currents
\begin{equation}
\{V_{\mu}(X(x)),\;V_{\nu}(X(y))\}=C_{\mu\nu}^{\lambda}V_{\lambda}%
(X(x))-\delta_{\mu\nu}\partial_{x}\delta(x-y). \label{eq96}%
\end{equation}
The more complicated case arise if both torsion are null tensors.

\subsection{Integrable string model with null torsion $C^{\mu\nu}_{\lambda}%
=0$}

To construct an integrable dynamical system we must to have a hierarchy of PBs
and\ we must to find a hierarchy of Hamiltonians through the bi-Hamiltonity
condition. The basic idea of the hydrodynamic approach given by Dubrovin,
Novikov \cite{Dub1}, \cite{Dub2} to integrable systems is a construction of
compatible local PB \cite{Mag} of an abelian currents from a pencil local PB
on the flat space of the currents and from the local PB on the curved space of
the currents. This approach was generalized by Ferapontov \cite{Fer1} and
Mokhov, Ferapontov \cite{Mok1} on the non-local PBs of hydrodynamic type. The
hydrodynamic type systems was considered by Tsarev \cite{Tsa}, Maltsev
\cite{Mal}, Ferapontov \cite{Fer2}, Mokhov \cite{Mok2}, \cite{Mok3}
\cite{Mok4}, Pavlov \cite{Pav3}, \cite{Pav4}, Maltsev, Novikov \cite{Mal}. The
Jacobi identity for compatible PB leads to the WDVV \cite{DW}, \cite{DVV}
associativity equation for the metric tensor of a curved space. Dubrovin
\cite{Dub3} shown that the WDVV equation is related to the Frobenius algebra
of a chiral currents. This equation is precisely the associativity condition
of the Frobenius algebra. Dubrovin and Zhang \cite{Dub4} gave many examples of
the solutions of the WDVV equation. We have applied the hydrodynamic approach
of the Dubrovin, Novikov and the Dubrovin solutions of the WDVV equation to
the description of the integrable string model in the background gravity field
with zero torsion \cite{Ger2}, \cite{Ger3}, \cite{Ger4}. Now let us to
consider string model with constant torsion.

\section{Integrable string model with the constant torsion}

We have considered the string model with constant torsion without
antisymmetric field $B_{ab}(X)$ in the light-cone gauge in the target space
\cite{Ger4}, \cite{Ger5}, \cite{Ger6}, \cite{Ger7}. This model coincides to
the principal chiral model on the compact simple Lie group. We can not
separate the dynamics on the right-mode only and the left mode for the
$\sigma$- model because of the initial chiral currents are not conserved
\[
\partial_{-}U_{\mu}=C_{\mu}^{\nu\lambda}U_{\nu}V_{\lambda}\neq0,\;\;\partial
_{+}V_{\mu}=C_{\mu}^{\nu\lambda}V_{\nu}U_{\lambda}\neq0.
\]
The correspondent charges are not Casimirs. These papers were motivated by the
papers \cite{Gol}, \cite{Are}, \cite{Eva}. Evans, Hassan, MacKay, Mountain
\cite{Eva} where they constructed the local invariant chiral currents
$C_{n}(U)$ (\ref{eq31}), (\ref{eq32}), as the scalar symmetric polynomials of
the initial chiral currents $U_{\mu}(X(x))$ of the $SU(n)$, $SO(n)$, $SP(n)$
principal chiral models. The constant torsion $C_{\nu\lambda}^{\mu}$ in the
commutation relations for the initial chiral currents $U_{\mu}(x)$
(\ref{eq23}) does not contribute to the equation of motion for the invariant
currents $C_{n}(U(X))$ through the total symmetrical functions $C_{n}(U)$. The
index $n$ of the invariant current $C_{n}$ is the power of the product of the
initial chiral currents $U_{\mu}$. However, the introduction of the
antisymmetric field $B_{ab}(X)$ to Lagrangian permits to obtain the equation
of motion for the initial chiral current $U_{\mu}(X(t,x))$.

Now, we can make use of the results section 3 to describe the string model of
the WZNW type with the constant torsion (see also papers \cite{Ger1},
\cite{Ger0}. The mathematical description of the equations of motion of the
invariant chiral and initial currents of the string model with the constant
torsions in this section coincides with the case of the WZNW model with the
constant torsion which was considered in Section 3.

\subsection{Integrable string models with $SU(2)$, $SO(3)$, $SP(2)$ constant
torsions and hydrodynamic chains}

The PB of Liouville coordinate $C_{2}(U(x))$ has the following forms:
\[
\{C_{2}(U(x)),C_{2}(U(y))\}=-2C_{2}(U(y))\partial_{y}\delta(y-x)+2C_{2}%
(U(x))\partial_{x}\delta(x-y).
\]
We will consider invariant chiral $C_{2}(U(x))$ as a local field on the
Riemmann space of the chiral currents. As the Hamiltonians we choose the
following functions
\[
H_{2(n+1)}=\frac{1}{2(n+1)}\int\limits_{0}^{2\pi}C_{2}^{n+1}%
(U(y))dy,\;n=0,1,...\infty.
\]
The equation of motion for the density of the first Casimir operator is as
follows
\[
\frac{\partial C_{2}}{\partial t_{2(n+1)}}-(2n+1)(C_{2})^{n}\frac{dC_{2}}%
{dx}=0.
\]
The equation for the currents $C_{2}^{n}=C_{2n}$ is following:
\[
\frac{\partial C_{2}^{n}}{\partial\tau_{n}}+(C_{2})^{n}\frac{dC_{2}^{n}}%
{dx}=0,\;\tau_{n}=-(2n+1)t_{2(n+1)}.
\]
This equation is inviscid Burgers equation \cite{Bur}. We will find the
solution in the form:
\[
C_{2}^{n}(\tau_{n},x)=\exp{(a+i(x-\tau_{n}C_{2}^{n}(\tau_{n},x)))}.
\]
following:
\[
C_{2}(t_{2(n+1)},x)=[\frac{i}{(2n+1)t_{2(n+1)}}W(-i(2n+1)t_{2(n+1)}%
e^{a+ix})]^{\frac{1}{n}}.
\]
Here function $W(t_{2(n+1)},x)$ is periodical Lambert function \cite{Lam}. The
equation of motion for the initial chiral current $U^{\mu}$ defined by the PB
(\ref{eq24}) and the Hamiltonian (\ref{eq40})
\[
\frac{\partial U_{\mu}}{\partial t_{2(n+1)}}=\frac{\partial}{\partial
x}[U_{\mu}(UU)^{n}]=C_{2}^{n}\frac{\partial}{\partial x}U_{\mu}+U_{\mu}%
\frac{\partial}{\partial x}C_{2}^{n},\;\mu=1,2,3.
\]
It is possible to rewrite this equation as the linear equation by using the
solution (\ref{eq45}) which diagonalize the equation (\ref{eq46})
\[
\frac{\partial U_{\mu}}{\partial t_{2(n+1)}}=\frac{\partial U_{\mu}}{\partial
x}f_{n}+U_{\mu}\frac{\partial}{\partial x}f_{n}%
\]
or as the linear nonhomogeneous equation
\[
\frac{\partial z^{\mu}}{\partial t_{2(n+1)}}=f(t_{n},x)\frac{\partial z^{\mu}%
}{\partial x}+\frac{\partial}{\partial x}f(t_{n},x),\;\;z^{\mu}=\ln U^{\mu
},\;f=C_{2}^{n},\;\frac{\partial z^{\mu}}{\partial x}=\frac{1}{U^{\mu}}%
\frac{\partial U^{\mu}}{\partial x},\;(not\;sum).
\]
The first example of the infinite dimensional hydrodynamic chains is based on
the invariant chiral currents $C_{2n}=(C_{2})^{n},\;n=1,2,...,\infty$ of the
string model with the $SU(2)$, $SO(3)$, $SP(2)$ constant torsions. The PB of
the different degrees of the invariant chiral currents $C_{2}^{n}(x)$,
$C_{2}^{m}(x)$ has form:
\[
\{C_{2}^{m}(x),C_{2}^{n}(y)\}=\frac{2nm(2m-1)}{n+m-1}C_{2}^{n+m-1}%
(x)\frac{\partial\delta(x-y)}{\partial x}-\frac{2nm(2n-1)}{n+m-1}C_{2}%
^{n+m-1}(y)\frac{\partial\delta(y-x)}{\partial y}.
\]
The equation of motion for invariant current $C_{2}^{m}$ with Hamiltonian
\[
H_{2n}=\frac{1}{2n}\int\limits_{0}^{2pi}C_{2n}(y)dy
\]
is
\[
\frac{\partial C_{2}^{m}}{\partial t_{2n}}=\frac{m(2n-1)}{m+n-1}\frac{\partial
C_{2}^{m+n-1}}{\partial x}.
\]
After the redefinition $C_{2}^{n}=C_{2n}=C_{p}$ we can obtain the standard
form of the hydrodynamic chain (\ref{eq52}).
\[
\{C_{p}(x),C_{q}(y)\}=\frac{pq(p-1)}{p+q-2}C_{p+q-2}(x)\frac{\partial
\delta(x-y)}{\partial x}-\frac{pq(q-1)}{p+q-2}C_{p+q-2}(y)\frac{\partial
\delta(y-x)}{\partial y}.
\]

The second example of the infinite dimensional chain is based on the invariant
chiral currents of the $\sigma-model$ with the $SU(\infty)$, $SO(\infty)$,
$SP(\infty)$ constant torsions. If the dimension of the matrix representation
$n$ is not ended $(n\rightarrow\infty)$ all the chiral currents are the
primitive ones. The PB in Liouville coordinates $C_{m}(x),\;m=2,3,...,\infty$
has the form (\ref{eq53}): (\ref{eq54}):
\[
\{C_{m}(x)C_{n}(y)\}=-W_{mn}(C(y))\frac{\partial}{\partial y}\delta
(y-x)+W_{nm}(C(x))\frac{\partial}{\partial x}\delta(x-y),
\]%
\[
W_{mn}(C(x))=\frac{mn(n-1)}{m+n-2}C_{m+n-2}(x).
\]
This PB satisfies to the skew-symmetric condition $\{C_{m}(x),C_{n}%
(y)\}=-\{C_{n}(y),C_{m}(x)\}$. The Jacobi identity imposes conditions on the
Hamiltonian function $W_{mn}(C(x))$ \cite{Dub1}, \cite{Dub2}, \cite{Mal1}
(look (\ref{eq53}):
\[
(W_{kp}+W_{pk})\frac{\partial W_{mn}}{\partial C_{k}}=(W_{km}+W_{mk}%
)\frac{\partial W_{pn}}{\partial C_{k}},\;\frac{dW_{kp}}{dx}\frac{\partial
W_{nm}}{\partial C_{k}}=\frac{dW_{km}}{dx}\frac{\partial W_{np}}{\partial
C_{k}}.
\]
The Jacobi identity satisfies for metric tensor $W_{mn}(C(x))$ (\ref{eq54}) .
The PB (\ref{eq53}) forms the algebra and can be rewritten as the PB of the
hydrodynamic type .
\[
\{C_{m}(x),C_{n}(y)\}=\frac{mn(n-1)}{m+n-2}\frac{dC_{m+n-2}(x)}{dx}%
\delta(x-y)+mnC_{m+n-2}(x)\frac{\partial}{\partial x}\delta(x-y).
\]
The algebra of charges $\int\limits_{0}^{2\pi}C_{n}(x)dx$ is the abelian
algebra. Let us choose the Casimir operators $C_{n}$ as the Hamiltonians
\[
H_{n}=\frac{1}{n}\int\limits_{0}^{2\pi}C_{n}(x)dx,\;n=2,3...\,.
\]
The equations of motion for the densities of Casimir operators are following
\[
\frac{\partial C_{m}(x)}{\partial t_{n}}=\frac{1}{n}\int\limits_{0}^{2\pi
}[-W_{mn}(C(y))\frac{\partial}{\partial y}\delta(y-x)+W_{nm}C((x))\frac
{\partial}{\partial x}\delta(x-y)]dy=\frac{m(n-1)}{m+n-2}\frac{\partial
}{\partial x}C_{m+n-2}.
\]
Thus the invariant chiral currents with the $SU(2)$, $SO(3)$, $SP(2)$ constant
torsion and the invariant chiral currents with the $SU(\infty)$, $SO(\infty)$,
$SP(\infty)$ constant torsion form the same infinite hydrodynamic chain
(\ref{eq52}), (\ref{eq53}), ({\ref{eq54}). These PBs (\ref{eq53}),
(\ref{eq54}) are a particular case of the $M$-brackets (Dorfman \cite{Dor} and
Kupershmidt \cite{Kup1}, \cite{Kup2} )for $M=2$ and describe the hydrodynamic
chains (see \cite{Pav1}, \cite{Pav2} and references therein). We can construct
the new nonlinear equations of motion for the initial chiral currents $U^{\mu
}$ using the flat $PB_{2}$ (\ref{eq24}) and Hamiltonians $H_{n}$ (\ref{eq57}),
where $C_{n}(x)$ defined by the equation (\ref{eq32}) for $SU(\infty)$ group:
\[
\frac{\partial U_{\mu}(x)}{\partial t_{n}}=\frac{1}{n}\int\limits_{0}^{2\pi
}dy\{U_{\mu}(x),C_{n}(U(y))\}_{2},
\]%
\[
\frac{\partial U_{\mu}(x)}{\partial t_{n}}=\frac{\partial}{\partial x}%
[d_{\nu_{1}\nu_{2}}^{k_{1}}d_{k_{1}\nu_{3}}^{k_{2}}...d_{\nu_{n-1}\mu
}^{k_{n-3}}U^{\nu_{1}}(x)...U^{\nu_{n-1}}(x)].
\]
As an example we consider $n=3$:
\[
\frac{\partial U_{\mu}}{\partial t_{3}}=\frac{\partial}{\partial x}(d_{\mu
\nu\lambda}U^{\nu}U^{\lambda}),\;\;\mu=1,2,...\infty.
\]
It is easy to see that this dynamical system is bi-Hamiltonian:
\[
\frac{\partial U_{\mu}(x)}{\partial t_{3}}=\frac{1}{3}\int\limits_{0}^{2\pi
}dy\{U_{\mu}(x),C_{3}(U(y))\}_{2}=\frac{1}{2}\int\limits_{0}^{2\pi}dy\{U_{\mu
}(x),C_{2}(U(y))\}_{3}.
\]
Here $PB_{3}$ has form:
\[
\{U_{\mu}(x),U_{\nu}(y)\}_{3}=2d_{\mu\nu\lambda}U^{\lambda}(x)\frac{\partial
}{{\partial x}}\delta(x-y).
\]
Let us remind that $d_{\mu\nu\lambda}$ are the symmetric structure constant of
the $SU(\infty)$ algebra in a matrix representation. This PB satisfies to
Jacobi identity for ($n\rightarrow\infty$)
\[
d_{\sigma\mu\nu}d_{\sigma\lambda\rho}+d_{\sigma\mu\lambda}d_{\sigma\nu\rho
}+d_{\sigma\mu\rho}d_{\sigma\nu\lambda}=\frac{1}{n}(\delta_{\mu\nu}%
\delta_{\lambda\rho}+\delta_{\mu\lambda}\delta_{\nu\rho}+\delta_{\nu\rho
}\delta_{\nu\lambda}).
\]
}

By the similar manner we can obtain the equation of motion for the chiral
currents of $SO(\infty)$, $SP(\infty)$:
\[
\frac{\partial U_{\mu}(x)}{\partial t_{n}}=\frac{\partial}{\partial x}%
[v_{\nu_{1}\nu_{2}\nu_{3}}^{k_{1}}...v_{\nu_{2n-2}\nu_{2n-1}\mu}^{k_{2n-3}%
}U^{\nu_{1}}...U^{\nu_{2n-1}}].
\]
As an example we consider $n=4$:
\[
\frac{\partial U_{\mu}}{\partial t_{4}}=\frac{\partial}{\partial x}(v_{\mu
\nu\lambda\rho}U^{\nu}U^{\lambda}U^{\rho}),\;\mu=1,2,...\infty.
\]
As final remark of this Section, we stress that the Casimir $C_{2}$ depend on
physical coordinates, having all the constraints conveniently solved.

\section{Concluding remarks}

We considered WZNW and string models as auxiliary problems to obtain
integrable equations of hydrodynamic type. We show that WZNW model coincided
to string mode for the transverse coordinateso of string and that these models
are integrable if the torsions are constant and they are the structure
constants of $SU(2),SO(3),SP(2)$ algebras. Also new integrable hydrodynamic
chains for $SU(\infty)$, $SO(\infty)$, $SP(\infty)$ are obtained.

\section{Acknowledgements}

Gershun V.D. should like to thank B.A. Dubrovin, O.I. Mokhov, M.V. Pavlov,
S.P. Tsarev, A.A. Zheltukhin for interest in his investigation and fruitful
discussions. Cirilo-Lombardo D.J. is very grateful to JINR\ and Bogoliubov
Laboratory of Theoretical Physics\ directorate for the hospitality and
financial support.


\section*{References}

\begin{thebibliography}{99}                                                                                               %


\bibitem {Wei}{\footnotesize S. Weinberg. Precise relations between the
spectra of vector and axial-vector mesons. \textit{Phys. Rev. Lett.}, 1967,
Vol. 18, p. 507-509. }

\bibitem {Sch}{\footnotesize J. Schwinger. Chiral dynamics. \textit{Phys.
Lett. B.}, 1967, Vol. 24, p. 473-476. }

\bibitem {WZ1}{\footnotesize J. Wess, B. Zumino. Lagrangian method for chiral
symmetries. \textit{Phys. Rev.}, 1967, Vol. 163, p. 1727-1735. }

\bibitem {Col}{\footnotesize S. Coleman, J. Wess, B. Zumino. Structure of
phenomenological lagrangians 1. \textit{Phys. Rev.}, 1969, Vol. 177, p.
2239-2247. }

\bibitem {Volk1}{\footnotesize D.V. Volkov. Phenomenological lagrangian
interection of goldstoun particles. \textit{preprint/ NAC Ukraine, Kiev, ITP},
1969, 51 pp. (in Russian). }

\bibitem {Volk}{\footnotesize D.V. Volkov. Phenomenological lagrangians.
\textit{Physics of elementary particles and atomic nuclei(PEPAN)}, 1973,
Vol.4, p. 3-41. (in Russian). }

\bibitem {VGT}{\footnotesize D.V. Volkov, V.D. Gershun, V.I. Tkach. Current
structure of phenomenological lagrangians. \textit{Theoretical and
mathematical physics}, 1970, Vol. 3, p.321-328. (in Russian). }

\bibitem {VZB}{\footnotesize D.V. Volkov, A.A. Zheltukhin, U.P. Bliokh.
Phenomenological lagrangian of spin waves. \textit{Phys. of the Solid State},
1971, Vol. 13, p. 1668-1678. (in Russian). }

\bibitem {WZ}{\footnotesize J. Wess, B. Zumino. Consequences of anomalous Ward
identities. \textit{Phys. Lett. B}, 1971, Vol. 37(1), p. 95--97. }

\bibitem {Wit1}{\footnotesize E. Witten. Global aspects of current algebra.
\textit{Nucl. Phys. B}, 1983, Vol. 223(2), p. 422--432. }

\bibitem {Wit2}{\footnotesize E. Witten. Non-abelian bosonization in two
dimensions. \textit{Commun. in Math. Phys.}, 1984, Vol. 92(4), p. 455--472. }

\bibitem {Nov1}{\footnotesize S.P. Novikov. Multivalued functions and
functionals. An analoque of the Morse theory. \textit{Sov. Math. Dokl.}, 1981,
Vol. 24, p. 222--226. }

\bibitem {Bra}{\footnotesize E. Braaten, T. Curtright, C. Zachos. Torsion and
geometrostasis in nonlinear sigma models. \textit{Nucl. Phys. B}, 1985, Vol.
260, p. 630--688. }

\bibitem {Bus}{\footnotesize T.H. Buscher. A symmetry of the string background
fields equations. \textit{Phys. Lett. B}, 1987, Vol. 194, p. 59-62. }

\bibitem {Tse}{\footnotesize A.A. Tseytlin. Conditions of Weyl invariance of
the two-dimensional sigma model from equations of stationarity of the "central
charge" action. \textit{Phys. Lett. B}, 1987, Vol. 194, p. 63-68. }

\bibitem {Mok6}{\footnotesize O.I. Mokhov. Symplectic forms on loop spaces of
manifolds and Riemmannian geometry. \textit{Functional Anal. Appl.}, 1990,
Vol. 24, p. 86-87. }

\bibitem {Mok5}{\footnotesize O.I. Mokhov. Symplectic and Poisson structures
on loop spasec of smooth maifolds and integrable systems. \textit{Uspekhi
Mathematical Surveys}, 1998, Vol. 53, p. 85-192; English translation in:
\textit{Russian Math. Serv.}, 1998, Vol. 53, p. 515-622. }

\bibitem {Kac}{\footnotesize V. Kac. Simple graded Lie algebras of finite
growth. \textit{Funkt. Anali. i ego Prilozhen.}, 1967, Vol. 1, p.
82-126.(English translation: \textit{Functional Anal. Appl.} 1967, Vol. 1, p.
328-372.) }

\bibitem {Moo}{\footnotesize R.V. Moody. Lie algebras associated with
generalized Cartan matrices. \textit{Bull. Amer, Math, Soc.}, 1967, Vol. 73,
p. 217-221. }

\bibitem {Zakh}{\footnotesize V.E. Zakharov, A.V. Mikhailov. Relativistically
invariant two-dimensional models in field theory integrable by inverse problem
technique. \textit{JETP}, 1974, Vol. 74, p. 1953-1973. (in Russian) }

\bibitem {Dol}{\footnotesize L. Dolan. Kac-Moody algebras is hidden symmetry
of chiral models. \textit{Phys. Rev. Lett.}, 1981, Vol. 47, p. 1371-1374. }

\bibitem {Non}{\footnotesize D.V. Volkov, V.D. Gershun, V.i. Tkach. About
nonlocal charge algebra in two-dimensional models. \textit{Ukrainian Journal
of Physics}, 1983, Vol. 28, p. 641-649. (in Russian) }

\bibitem {Gol}{\footnotesize Y.Y. Goldschmidt and E. Witten. Conservation laws
in some two-dimensional models. \textit{Phys. Lett. B}, 1980, Vol. 91, p.
392--396. }

\bibitem {Are}{\footnotesize I.Ya. Aref'eva, P.P. Kulish, E.R. Nissimov and
S.J. Pacheva. Infinite set of conservation laws of the quantum chiral field in
2D spacetime. preprint {LOMI E-I-1978}, 1978, unpublished. }

\bibitem {Eva}{\footnotesize J.M. Evans, M. Hassan, N.J. MacKay, A.J.
Mountain. Local conserved charges in principal chiral models. \textit{Nucl.
Phys. B}, 1999, Vol. 561, p. 385--412; \eprint{hep-th/9902008}. }

\bibitem {Mou}{\footnotesize A.J. Mountain. Invariant tensors and Casimir
operators for simple compact groups. \eprint{physics/9802012}. }

\bibitem {Az}{\footnotesize J. A. de Azcarraga, A. J. Macfarlane, A. J.
Mountain and J. C. Perez Bueno. Invariant tensors for simple groups.
textit{Nucl. Phys. B}, 1998, Vol. 510, P. 657-687; \eprint{physics/9706006}. }

\bibitem {Kle}{\footnotesize A. Klein, Invariant operators of the unimodular
group in n dimensions. \textit{J. Math. Phys.}, 1963, Vol. 4, p. 1283--1284. }

\bibitem {Ras}{\footnotesize M.A. Rashid, Saifuddin. Identity satified by thr
d-type coefficients 0f SU(n). \textit{J. Math. Phys.}, 1973, Vol. 14,
p.630631. }

\bibitem {Sud}{\footnotesize A. Sudbery, Ph.D. Thesis. \textit{Cambridge
Univ.}, 1970; Computer-friendly d-tensor identities for SU(n), \textit{J.
Phys. A}, 1990, Vol. 23(15), p. L705--L710. }

\bibitem {Dub1}{\footnotesize B.A. Dubrovin, S.P. Novikov. Hamiltonian
formalism of one-dimensional systems of hydrodynamical type and the
Bogolyubov-Whitham averaging method, \textit{Soviet Math. Dokl.}, 1983, Vol.
27, p. 665--669. }

\bibitem {Dub2}{\footnotesize B.A. Dubrovin, S.P. Novikov. Hydrodynamics of
weakly deformed soliton lattices, \textit{\ Russian Math. Surveyes}, 1989,
Vol. 44, p. 35--124. }

\bibitem {Ger2}{\footnotesize V.D. Gershun. Integrable string models of
hydrodynamic type, \textit{J. Kharkov Univ., Phys. Ser.Nucl. Part.}, 2005,
Vol. 657, p. 109--113. }

\bibitem {Ger3}{\footnotesize V.D. Gershun. Integrable string models and
sigma-models of hydrodynamic type in terms of invariant chiral currents.
\textit{Prob. Atom. Sci. Technol.}, 2007, Vol. 3(1), p. 16--21. }

\bibitem {Ger4}{\footnotesize V.D. Gershun. Integrable string models in terms
of chiral invariants of $SU(n)$, $SO(n)$, $SP(n)$ groups,
\href{http://www.emis.de/journals/SIGMA/2008/041/}{\textit{\ SIGMA}}, 2008,
Vol. 4, 041, 16 pp.; \eprint{nlin. SI/0805.0656} }

\bibitem {Pav1}{\footnotesize M.V. Pavlov. Integrable hydrodynamic chains,
\textit{J. Math. Phys.}, 2003, Vol. 9, p. 4134--4156. }

\bibitem {Pav2}{\footnotesize M.V. Pavlov. Hydrodynamic chains and a
classification of their Poisson Brackets, \eprint{nlin. SI/ 0603056}. }

\bibitem {Dor}{\footnotesize I.Ya. Dorfman. Dirac structures and integrability
of nonlinear evolution equations. \textit{Nonlinear Science: Theory and
Applications}, John Wiley and Sons, New York,. 1993, 176 pp. }

\bibitem {Kup1}{\footnotesize B.A. Kupershmidt. Deformations of integrable
systems. \textit{Proc. Roy. Irish Acad. Sect. A}, 1983, Vol. 83, p. 45-74. }

\bibitem {Kup2}{\footnotesize B.A. Kupershmidt, Yu.I. Manin. Long wave
equations with a free surface. II. The Hamiltonian structure and the higher
equations. \textit{Func. Anal. Appl.}, 1978, Vol.12, p. 25-37. }

\bibitem {Gre}{\footnotesize M.B. Green, J.H. Schwarz, E. Witten. Superstring
theory, \textit{Superstring theory}, Cambridge University Press, Cambridge,
1987, Vol.1, 469 pp. }

\bibitem {Rud}{\footnotesize R.E. Rudd. Light-cone gauge quantization of 2D
sigma models, \textit{Nucl. Phys. B}, 1994, Vol. 427, p. 81--110;
\eprint{hep-th/9402106}. }

\bibitem {Mag}{\footnotesize F. Magri. A simple model of the integrable
Hamiltonian equation, \textit{J. Math. Phys.}, 1978, Vol. 19, p. 1156--1162. }

\bibitem {DW}{\footnotesize R. Dijkgraaf, E. Witten. Mean field theory,
topological field theory and multi-matrix models \textit{Nucl. Phys. B}, 1990,
Vol. 342, p. 486--522. }

\bibitem {DVV}{\footnotesize R. Dijkgraaf, E.Verlinde, H. Verlinde.
Topological strings in d
$<$
1, \textit{Nucl. Phys. B}, 1991, Vol. 352, p. 59--86. }

\bibitem {Fer1}{\footnotesize E.V. Ferapontov. Differential geometry of
nonlocal Hamiltonian of hydrodynamical type. \textit{Functional Anal. Appl.},
1991, Vol. 25, p. 195--204. }

\bibitem {Mok1}{\footnotesize O.I. Mokhov, E.V. Ferapontov. Nonlocal
Hamiltonian operators of hydrodynamical type related to metric of constant
curvature, \textit{\ Russian Math. Surveys} 1990, Vol. 45, p. 218--219. }

\bibitem {Mok2}{\footnotesize O.I. Mokhov. Symplectic and Poisson structures
on loop spaces of smooth manifolds and integrable systems. \textit{Russian
Math. Surveys}, 1990, Vol. 53, p. 515--622. }

\bibitem {Tsa}{\footnotesize S.P. Tsarev. On poisson brackets and
one-dimensional Hamiltonian systems of hydrodynamical type. \textit{Soviet
Math. Dokl.}, 1985, Vol. 31, p. 488--491. }

\bibitem {Mal}Andrei Ya. Maltsev, On the compatible weakly-nonlocal Poisson
brackets of Hydrodynamic Type, Intern. Journ. of Math. and Math. Sci. 32:10
(2002), 587-614.{\footnotesize ; \eprint{nlin.SI/0111015}. }

\bibitem {Fer2}{\footnotesize E.V. Ferapontov, Nonlocal Hamiltonian operators
of hydrodynamical type: differential geometry and applications. \textit{Amer.
Math. Soc. Transl.(2)}, 1995, Vol. 170,p. 33--58. }

\bibitem {Mok3}{\footnotesize O.I. Mokhov. Liouville form for compatible
nonlocal brackets of hydrodynamical type and integrable hierarchies.
\textit{Funcsional. Analiz i Prilozen.}, 2003, Vol. 37, p. 28--40;
\eprint{math.DG/0201223}. }

\bibitem {Mok4}{\footnotesize O.I. Mokhov. Compatible Dubrovin-Novikov
Hamiltonian operators, Lie derivative and integrable systems of hydrodynamical
type. \textit{\ TMF}, 2002, Vol. 133,p. 279--288; \eprint{math.DG/0201281}. }

\bibitem {Pav3}{\footnotesize M.V. Pavlov. Elliptic coordinates and
multi-Hamiltonian structure of the Whitham equations. \textit{Russian Acad.
Sci. Doklady Math.}, 1995, Vol. 50,p. 220--223. }

\bibitem {Pav4}{\footnotesize M.V. Pavlov. Integrability of Egorov
hydrodynamic type systems. \textit{Teor. and Math. Fizika}, 2007, Vol. 150,p.
263--285; \eprint{nlin/0606017}. }

\bibitem {Mal1}{\footnotesize A.Ya. Maltsev, and S.P. Novikov. On the local
systems Hamiltonian in the weakly nonlocal Poisson brackets. \textit{Physica
D}, 2001, Vol. 156, p. 53-80. \eprint {nlin.SI/0006030}. }

\bibitem {Dub3}{\footnotesize B.A. Dubrovin. Geometry of 2D topological field
theories. \textit{\ Lecture Notes in Physics}, 1996, Vol. 1620, p. 120--348;
\eprint{hep-th/9407018}. }

\bibitem {Dub4}{\footnotesize Boris Dubrovin, Youjin Zhang. Bihamiltonian
hierarchies in 2D topological field theory at one-loop approximation.
\textit{Commun. Math. Phys.}, 1998, Vol. 198, p. 311--361. }

\bibitem {Ger5}{\footnotesize V.D. Gershun. Integrable string models of
hydrodynamical type in terms of chiral currents. \textit{Proceedings of
International Workshop "Supersymmetries and Quantum Symmetries"
(SQS'05)--Dubna, Russia, July 27-31}, 2005, Dubna: JINR, 2006, p. 219--230. }

\bibitem {Ger6}{\footnotesize V.D. Gershun. Integrable string models and
sigma--models of hydrodynamic type in terms of chiral currents.
\textit{Proceedings of the 5th International Conference
Bolya--Gauss--Lobachevsky (BGL-5)"Non--Euclidean Geometry in Modern Physics",
Minsk, Belorus, October 10-13}, 2006, Minsk: In-t Phys., 2006, p. 248--257. }

\bibitem {Ger7}{\footnotesize V.D. Gershun. Integrable string models with
constant torsion in terms of chiral invarians of $SU(n)$, $SO(n)$, $SP(n)$
groups. \textit{Phys. Atom. Nucl.}, 2010, Vol. 73, p. 304--310;
\eprint{hep-th/ 0902.0542}. }

\bibitem {Ger1}{\footnotesize V.D. Gershun. Integrable string models with
constant $SU(3)$ torsion, \textit{Phys. Part. and Nucl. Lett.}, 2011, Vol. 8,
p. 293--298. }

\bibitem {Ger0}{\footnotesize V.D. Gershun. Integrable WZNW models and string
models of WZNW model type with constant SU(2) torsion. \textit{Prob. Atom.
Sci. Technol.}, 2012, Vol. 1(77), p. 337---21. }

\bibitem {Bur}{\footnotesize J.M. Burgers. The nonlinear diffusion equation.
\textit{New York: Springer-Verlag, LLC }, 2002, 188 pp. }

\bibitem {Lam}{\footnotesize J.H. Lambert. Observationes variae in mathesin
puram. \textit{Acta Helvet. Phys. Math. Botan. Med.}, 1758, Band III, p.
128-168. (facsimile (http://www.kuttaka.org/~JHL/L1758c.pdf)). }

\bibitem {Ger9}{\footnotesize V.D. Gershun. Integrable string models of WZNW
model type with constant $SU(2)$, $SO(3)$, $SP(2)$ and $SU(3)$ torsions and
hydrodynamic chains. \textit{Phys. of Part. and Nucl.}, 2012, Vol. 43, p.
659-662. }

\bibitem {Is}A.P. Isaev, Fermionic String Model In Lie Group Spaces,
Teor.Mat.Fiz. 71, 1987, 395-409 (Theor.Math.Phys. 71, 1987, 616-626).

\bibitem {Is1}A.P. Isaev, E.A. Ivanov, Green-schwarz Superstring As An
Asymmetric Chiral Field Sigma Model, Teor.Mat.Fiz. 81, 1989, 420-433
(Theor.Math.Phys. 81 (1990) 1304-1313).

\bibitem {Is2}A.P. Isaev, E.A. Ivanov, Nonabelian N=2 superstrings,Trieste,
ICTP preprint IC-90-97 ,1990 (unpublished); e-Print: arXiv:0912.2200 [hep-th].

\bibitem {Is3}A.P. Isaev, E.A. Ivanov, Nonabelian N=2 Superstrings:
Hamiltonian Structure, Trieste, ICTP preprint IC-91-86, 1991, unpublished;
e-Print: arXiv:0912.2204 [hep-th]
\end{thebibliography}
\end{document}